%% file: jdl.tex
\documentstyle [epsfig,epsf,rotate]{mn}

\include{mnmacros}

\begin{document}

\title[The tidal disruption of protoplanetary discs]
{The tidal disruption of protoplanetary accretion discs}

\author[J.D. Larwood]
{John D. Larwood  \\ 
Astronomy Unit, School of Mathematical Sciences, Queen Mary \&
Westfield College, Mile End Road, London E1 4NS}

\date{Accepted 1997 May 30}

\volume{000}
\pagerange{\pageref{firstpage}--\pageref{lastpage}}
\pubyear{0000}

\maketitle

\label{firstpage}

\begin{abstract}

\noindent In this paper we revisit the problem of the tidal
interaction occuring between a protostellar accretion disc
and a secondary point mass following a parabolic trajectory.
We model the disc response analytically and we compare our
results with three-dimensional SPH simulations.

\noindent Inviscid as well as viscous hydrodynamics is considered.
We show that in a viscous system the response derived
from inviscid considerations is predominant even for the highest
estimates of an anomalous disc shear viscosity. The angular momentum lost
from the disc during the encounter is derived from linear theory, for
distant fly-bys, as well as the changes to the disc orientation expected
in non-coplanar encounters.  

\noindent It is shown that the target discs can
become warped and precess by a small amount during non-coplanar
encounters. This small precession is shown to give rise to a relative tilt
of the disc which is always more important for determining its
final orientation than is the change to the orbital inclination.

\noindent We discuss the implications of our results for protostellar
accretion discs and planetary systems.

\end{abstract}


\section{Introduction}

\noindent Low mass stars are known to form in compact groups within
the dense cores of giant molecular clouds. Recent observations of
nearby star-forming regions have revealed that young stellar objects
(YSOs) have a binary frequency in excess of that found for field stars
and roughly half of the current sample are associated with optically thick, 
geometrically thin, circumstellar accretion discs (see recent
reviews by Mathieu 1994, Lin \& Papaloizou 1996 and references
therein). If the mean free path in a star-forming core is sufficiently
small so as to allow close encounters between YSOs on a timescale
shorter than the
lifetime of a protostellar disc then the tidal interaction between a
secondary object and the disc of the primary may be a dynamically
significant event in the disc's history. 

\noindent The possibility that YSOs might tidally interact through
their circumstellar discs has lead to recent speculation that
binaries may form as the products of a capture process
involving star-disc encounters (Larson 1990). However it was quickly
shown that even for the most compact star-forming environments
known, disc penetrating encounters leading to capture would be rare
(Clarke \& Pringle 1991), with the probability of such an encounter
being approximately one tenth over a disc lifetime. Also observational
data implies that most YSO binaries have components of similar ages
(Hartigan, Strom \& Strom 1994, Brandner \& Zinnecker 1997), which supports
local formation of the components rather than capture.
We note however that more distant encounters are near
certainties for protostellar accretion discs.

\noindent These kinds of star-disc
interactions have been extensively studied in the context of star
formation. Heller (1993) investigated numerically with SPH the rate of
disc tilting resulting from non-coplanar disc penetrating encounters.
Clarke \& Pringle (1993) investigated numerically, but with a
sticky particle code, the affect of a disc penetrating encounter on
the distribution of disc material. Ostriker (1994) applied linear
perturbation theory to an inviscid hydrodynamical model, using
a spherical harmonic expansion of the tidal potential for distant
encounters, to derive an asymptotic expression for the angular
momentum lost from the disc in the encounter. Korykansky \&
Papaloizou (1995) used a Fourier expansion in azimuthal modes,
for the tidal potential, and calculated the angular momentum exchange
without asymptotic assumptions. Hall, Clarke \& Pringle (1996)
carried out a numerical investigation of the disc response
using a reduced three-body method with non-interacting particles.

\noindent Due to the complexity of the tidal problem the analytical
results presently available in the literature, although extensive,
are also unweildy and the numerical studies are chiefly qualitative.
In this paper we present analytical and numerical (SPH)
calculations of the disc response to a parabolic encounter.
We present a simplified analysis isolating the most significant parts
of the response which have been identified in the works listed above.
We obtain explicit expressions for the angular momentum exhange (and
other properties of the response) and we test these results directly
against the simulations.

\noindent In Section $2$ we give some of the equations basic to our
subsequent considerations. In Section $3$ we review concepts
of the linear disc response and perform a linear mode analysis
of the fluid equations. In Section $4$ we derive the linear
disc response for distant coplanar encounters and in Section $5$ we
extend this to consideration of the response for distant non-coplanar
encounters. Our numerical method is outlined in Section $6$ and in
Section $7$ we
present our numerical results. In Section $8$ we summarise our findings
and discuss our results within the context of recent observations.

\section{Basic Equations}

\subsection{The accretion disc model}

\noindent We consider a non--self-gravitating gas-dynamical model for
the accretion disc.
Processes that may give rise to internal heat
generation or transport are not considered. Instead we adopt a simple
polytropic relationship between gas pressure, $P$, and density, $\rho$, giving
a constitutive equation of state:

\begin{equation}
P= K \rho^{1+1/n},
\label{stat}
\end{equation}

\noindent where $n$ is the polytropic index and $K$ is
the polytropic constant. The associated barotropic sound speed,
$c_{\rm s}$, is given by

\begin{displaymath}
c_{{\rm s}}^2 = \frac{{\rm d}P}{{\rm d}\rho}.
\end{displaymath}

\noindent The constancy of $K$ in this model requires that we assume
any dissipated energy to be lost from the system. This is equivalent
to assuming an efficient cooling mechanism operating in the disc.

\subsubsection{Equilibrium structure}

Referring to a set of cylindrical coordinates $(r,\phi,z)$ with
the $z$-axis coincident with
the rotation axis of the disc, the standard equation of vertical
hydrostatic equilibrium in the thin disc approximation is: 

\begin{equation}
{1 \over \rho}{\upartial P \over \upartial z} = -\Omega^2 z ,
\label{vert}
\end{equation}

\noindent where $\Omega= \Omega(r)$ is the Keplerian angular velocity of disc
material in an initially axisymmetric disc.
For a disc satisfying the polytropic equation of state (\ref{stat})
integration of equation (\ref{vert}) gives

\begin{displaymath}
\rho (r,z) = \left[{\Omega^2H^2\over 2K(n+1)}\right]^n(1-z^2/H^2)^n,
\end{displaymath} 

\noindent where $H=H(r)$ is the total vertical semi--thickness for
which $\rho(z=\pm H)=0$. We may now compute the sound speed
in the disc in terms of the midplane value, $\bar{c}_{{\rm s}}$:

\begin{displaymath}
c_{{\rm s}}^2 = \bar{c}_{{\rm s}}^2 (1-z^2/H^2) ,
\end{displaymath}

\noindent the midplane value being given by
$\bar{c}_{{\rm s}}^2 = \Omega^2 H^2 / 2n$.

\noindent Following a thin disc approximation we shall also consider
the disc surface density $\Sigma$, defined by:

\begin{displaymath}
\Sigma \equiv \int_{-H}^{+H} \rho {\rm d}z
= {\cal C}_{n} H {\left[ \frac{\Omega^{2} H^{2}}{2K(n+1)} \right]}^n .
\end{displaymath}

\noindent The constant ${\cal C}_{n}$ is
given by the {\em beta function} $\beta(\frac{1}{2}, n+1)$.

\noindent We do not consider viscous forces explicitly in the equation
of motion, but where required we make use of the standard
$\alpha$--prescription due to Shakura \& Sunyaev (1973). In this
prescription the usual coefficient of kinematic shear viscosity,
$\nu$, is written as $\nu = \alpha c_{{\rm s}} H$. In this case $\alpha$
is the dimensionless parameter to be determined. The viscous
evolution timescale is $\sim r^{2}/\nu$, being much larger than the dynamical 
timescale, $\Omega^{-1}$, which is also the timescale on which
vertical hydrostatic equilibrium can be established.

\subsection{The secondary}

\noindent Initially we shall consider the case in which the disc midplane
and the orbital plane of the secondary coincide. Our results are then
extended to include the case in which the two planes are mis-aligned.
We take the primary and the secondary to be point masses.
Neglecting any energy exchange phenomena that might occur between
the secondary and the disc the energy per unit mass of the secondary
referred to the non-inertial frame centred on the primary is written:

\begin{equation}
E= \frac{{\dot{\sigma}}^{2} }{2} - 
\frac{ G(M_{{\rm p}} + M_{{\rm s}}) }{\sigma}
+ \frac{\sigma^{2}{\dot{\theta}}^2}{2} .
\label{energy}
\end{equation}

\noindent In a coplanar configuration the position of the secondary at
time $t$ is ${\bmath D}(t)=[\sigma(t), \theta(t), 0]$. $M_{{\rm p}}$
is the mass of
the primary, $M_{{\rm s}}$ is the mass of the secondary and we
denote with a dot
the derivative with respect to time. For a parabolic
trajectory we set $E=0$ and integrate (\ref{energy}) with respect
to time. Referring to a set of Cartesian coordinates $(x,y,z)$ we can choose
the $y$-axis to coincide with the longitude of pericentre. In this
case the position of the secondary as a function of time is given
parametrically:

\begin{displaymath} 
{\bmath D} = q[4p, (1-4p^{2}), 0] ,
\end{displaymath}

\noindent in which

\begin{displaymath}
p = \sinh \left[ \frac{1}{3} \sinh^{-1}
\left( \frac{3}{4}\omega_{{\rm o}}t \right) \right] .
\end{displaymath}

\noindent The distance of closest approach of the secondary to the primary is
denoted by $q$ and the magnitude of the angular velocity of the secondary at
pericentre by $\omega_{{\rm o}}$:

\begin{displaymath}
\omega_{\rm o}= \frac{\sqrt{2G(M_{\rm p} + M_{\rm s})}}{q^{3/2}} .
\end{displaymath}

\noindent The natural origin of time corresponds to the instant
of pericentre passage, i.e. $D \equiv |{\bmath D}|= q$ at $t=0$.

\subsection{Tidal potential for distant encounters}

\noindent We consider a geometrically thin disc, initially
in a state of vertical hydrostatic equilibrium and radial centrifugal
equilibrium, governed by the central potential ${\Psi}_{{\rm 0}}$
alone. This initial
state is then perturbed from equilibrium by the tidal forcing due to a
secondary point mass encountering the primary/disc on a parabolic
prograde orbit.
The part of the total potential that is due to the secondary,
${\Psi}^{\prime}$, is considered to drive small perturbations of the disc's
equilibrium structure. Within the context of a fluid model we can then
apply a linear perturbation
analysis of the fluid equations. This will always be valid if the
affect of the tide is sufficiently weak, and amounts to considering
the distance of closest approach $q$ to be large when compared to
the outer radius of the disc $R_{{\rm o}}$. The total potential is
$\Psi = {\Psi}_{{\rm 0}} + {\Psi}^{\prime}$,
where the potential due to the primary acting at a point with
position vector ${\bmath r}$ is given by

\begin{displaymath}
{\Psi}_{\rm 0} = - \frac{GM_{\rm p}}{\mid \bmath{r} \mid} 
\end{displaymath}

\noindent and the potential due to the secondary, referred to the
origin which is fixed to the primary, is given by

\begin{equation}
{\Psi}^{\prime} = -\frac{GM_{{\rm s}}}{\mid \bmath{r} - \bmath{D} \mid} +
{GM_{{\rm s}}{\bmath r} \cdot {\bmath D}\over D ^3}.
\label{secpot}
\end{equation}

\noindent Expanding up to second order in terms proportional to $r/D$, and
ignoring terms proportional to $(z/r)^{2}$,
we may then write the part of the total potential due to the secondary
as (Papaloizou \& Terquem 1995)

\begin{equation}
{\Psi}^{\prime} = -\frac{GM_{{\rm s}}}{D}
\left[ 1 - \frac{{\mid \bmath{r} \mid}^2}{2D^2}
+ \frac{3{({\bmath r} \cdot {\bmath D})}^2}{2D^4} \right].
\label{psis}
\end{equation}

\section{The disc response}

\subsection{Wave generation and propagation}

\noindent For tightly wound spiral density waves propagating with a radial
wave number of magnitude $|k_{r}|$ in a geometrically thin
self-gravitating disc the standard dispersion relation
(Lin \& Shu 1964) is written:

\begin{equation}
m^{2}{(\Omega-\Omega_{\rm p})}^{2} = \kappa^{2}-2\upi G \Sigma |k_{r}|
+ {|k_{r}|}^{2} c_{\rm s}^{2} . 
\label{disp1}
\end{equation}

\noindent In the non--self-gravitating limit this may be expressed as

\begin{equation}
|k_{r}| = \frac{\kappa}{c_{{\rm s}}} \sqrt{s^{2}-1} ,
\label{disp2}
\end{equation}

\noindent where $s$ (Binney \& Tremaine 1987) gives the ratio of the forcing
frequency in a frame rotating with pattern speed $\Omega_{{\rm p}}$ to
the natural
radial frequency $\kappa$ (also called the epicyclic frequency):

\begin{displaymath}
s= \frac{m(\Omega - \Omega_{\rm p})}{\kappa}
\end{displaymath}

\noindent and

\begin{displaymath}
\kappa^2 \equiv \frac{1}{r^3}{{\rm d} \over {\rm d}r}(r^4{\Omega^2}) .
\end{displaymath}

\noindent The positive integer $m$ is the azimuthal mode number.
Resonances occur at
$s= 0, +1, -1$ corresponding to the corotation resonance and the inner and
outer Lindblad resonances respectively. When a fluid disc is subject to a
perturbing force which is stationary in a frame rotating with a
constant angular frequency $\Omega_{{\rm p}}$, a stationary disturbance is set up
in that frame, the disturbance having the same order of azimuthal symmetry
as the perturbing force.

\noindent The radial wave number given by (\ref{disp2}) indicates short
wavelength spiral density waves which may be either inward or outward
propagating. Furthermore, standard results from the analysis of the
solutions of the dispersion relation (\ref{disp1}) are
that for non--self-gravitating discs there exists an evanescent zone
about the corotation radius which occupies the region between the
inner and outer Lindblad resonances. Inside the evanescent zone wave
propagation is formally disallowed. We can further deduce that the spiral
density waves become increasingly tightly wound as they travel away
from the Lindblad resonances.  

\noindent In the tidal problem there is a torque exerted on the disc by the
secondary. The torque principally interacts with the disc through the Lindblad
resonances (Goldreich \& Tremaine 1979), provided that they are either
in, or lie sufficiently near
to the disc (Lin \& Papaloizou 1993). The applied torque excites short
trailing waves at the Lindblad resonances which propagate away
from the evanescent zone with a wavelength $\propto H$. As long as
the waves remain linear then they
propagate with a conserved wave action. Inward propagating waves then become
increasingly tightly wound and increase in amplitude until non-linear
damping takes place and the angular momentum they transport is deposited
in the disc. This is a well known process that may lead to induced accretion
near the disc centre. The net angular momentum
transport due to the corotation resonance is much smaller
and depends on the ratio $4\Omega\kappa^{2}/\Sigma$ (Korykansky \&
Papaloizou 1995), evaluated at the resonant location.

\noindent Consideration of the full problem for parabolic fly-bys
requires analysis of the disc response to an infinite spectrum of
forcing frequencies. Recent work in this area (Ostriker 1994, Korykansky \&
Papaloizou 1995) has shown that
a single component of the response predominates: the part of
the tidal potential with $m=2$ azimuthal symmetry applied at pericentre.

\noindent Assuming that the secondary-disc interaction occurs as an impulse
near pericentre we shall treat the disc response as resulting from a
non-resonant interaction. In this time-independent calculation the
secondary is assumed to be fixed at pericentre so that the
torque exerted on the disc can be calculated in a simple way.
In practice the interaction has a finite width in the sense that a
resonant perturbation may be set up at some time before pericentre
passage. As a result a phase lag between the density response
of the disc and the forcing potential is set up,
allowing a non-vanishing net tidal torque to be applied to the disc.

\noindent We shall determine the disc response with the secondary
fixed at its
pericentre position using a linear mode analysis of the fluid equations and
apply the calculated torque over the estimated time interval
of the interaction.
Korykansky \& Papaloizou (1995) report that the
interaction interval they observe in their numerical models is
about one orbital period at the outer edge of the disc, centred at
the instant of pericentre passage.

\subsection{Linear mode analysis}

\noindent The vertically averaged inviscid fluid equations (see Papaloizou \&
Lin 1995b, for a recent review) yield radial
and azimuthal components of the momentum equation:

\begin{displaymath}
\frac{\upartial V_{{r}}}{\upartial t} + 
V_{{r}} \frac{\upartial V_{{r}}}{\upartial r} +
\frac{V_{\phi}}{r}\frac{\upartial V_{{r}}}{\upartial \phi} -
\frac{V_{\phi}^2}{r} =
-\frac{1}{\Sigma}\frac{\upartial P}{\upartial r} -
\frac{\upartial \Psi}{\upartial r} ,
\end{displaymath}

\begin{displaymath}
\frac{\upartial V_{\phi}}{\upartial t} + 
V_{{r}} \frac{\upartial V_{\phi}}{\upartial r} +
\frac{V_{\phi}}{r} \frac{\upartial V_{\phi}}{\upartial \phi} +
\frac{V_{\phi} V_{{r}}}{r} =
-\frac{1}{\Sigma r} \frac{\upartial P}{\upartial \phi} -
\frac{1}{r} \frac{\upartial \Psi}{\upartial \phi} , 
\end{displaymath}

\noindent and the continuity equation:

\begin{displaymath}
\frac{\upartial \Sigma}{\upartial t} + 
\frac{1}{r} \frac{\upartial}{\upartial r} (\Sigma r V_{{r}}) +
\frac{1}{r^2} \frac{\upartial}{\upartial \phi} (\Sigma r V_{\phi})
=0 .
\end{displaymath}

\noindent Note that within the context of the vertical averaging procedure
physical quantities are formally replaced by their
vertically averaged analogs. The radial and azimuthal velocity components
are denoted by $V_{r}$ and $V_\phi$ respectively. 

\noindent We proceed by making a standard linear
analysis of the vertically averaged fluid equations such that
quantities are perturbed by a small amount from their
equilibrium states (e.g. $\Sigma \rightarrow \Sigma_{{\rm 0}}
+ \Sigma^{\prime}$ in standard notation). Then subtracting the
equations for equilibrium and neglecting quantities of second and
higher orders in the perturbation one
obtains the fluid equations linear in perturbed quantities:

\begin{displaymath}
\frac{\upartial V_{{r}}^{\prime}}{\upartial t} + 
\Omega \frac{\upartial V_{{r}}^{\prime}}{\upartial \phi} -
2 \Omega V_{\phi}^{\prime}=
-\frac{\upartial}{\upartial r} 
\left( \frac{P^{\prime}}{\Sigma_{{\rm 0}}} + \Psi^{\prime} \right) ,
\end{displaymath}

\begin{displaymath}
\frac{\upartial V_{\phi}^{\prime}}{\upartial t} + 
\Omega \frac{\upartial V_{\phi}^{\prime}}{\upartial \phi} +
\frac{V_{{r}}^{\prime}}{r}
\left[ \frac{\upartial}{\upartial r} (r^{2} \Omega) \right] =
-\frac{1}{r} \frac{\upartial}{\upartial \phi} 
\left( \frac{P^{\prime}}{\Sigma_{{\rm 0}}} + \Psi^{\prime} \right) ,
\end{displaymath}

\noindent and

\begin{displaymath}
\frac{\upartial \Sigma^{\prime}}{\upartial t} + 
\frac{1}{r} \frac{\upartial}{\upartial r}
(\Sigma_{{\rm 0}} r V_{{r}}^{\prime}) +
\Omega \frac{\upartial \Sigma^{\prime}}{\upartial \phi}+
\frac{\Sigma_{{\rm 0}}}{r} \frac{\upartial V_{\phi}^{\prime}}{\upartial \phi}
=0 .
\end{displaymath}
   
\noindent These equations describe the linear response of the model.
We analyse for azimuthal modes such that we make replacements like
$\Sigma^{\prime} \rightarrow \Sigma_{{\rm a}}(r)\exp(m{\rm i}\phi)$, noting
that physically meaningful results are obtained by taking the real
parts. The complex amplitude functions for the time-independent forced
velocity components are (Goldreich \& Tremaine 1979):  

\begin{displaymath}
V_{\phi {\rm a}}= \frac{1}{\Delta} \left[
\frac{m^{2}\Omega}{r} +
\frac{\kappa^{2}}{2\Omega} \frac{\upartial}{\upartial r} \right]
\Psi_{\rm a}
\end{displaymath}

\noindent and

\begin{displaymath}
V_{r{\rm a}}= \frac{-{\rm i}}{\Delta} \left[
\frac{2m\Omega}{r} +
m\Omega \frac{\upartial}{\upartial r} \right]
\Psi_{{\rm a}} ,
\end{displaymath}

\noindent where
$\Delta \equiv \kappa^{2} - m^{2}\Omega^2$. Additionally we assume a thin
disc (i.e. $H/r$ is a small quantity) for the duration of the
encounter so that we can
neglect the pressure perturbation in comparison to the perturbation
to the potential. The perturbation to the surface density is then determined
from the perturbed continuity equation. In terms of the velocity 
perturbations,

\begin{displaymath}
\Sigma_{{\rm a}}= \frac{\rm i}{m\Omega} \left[
m{\rm i}\frac{\Sigma_{{\rm 0}}V_{\phi {\rm a}}}{r} +
\frac{1}{r}\frac{\upartial}{\upartial r}
(\Sigma_{{\rm 0}}rV_{r{\rm a}}) \right] .  
\end{displaymath}

\noindent We note that the denominator, $\Delta$, can vanish
for $m=1$ azimuthal modes, since for strictly Keplerian rotation
$\kappa=\Omega$. This leads to a singularity in the disc
response equation which is usually dealt with by employing
a dissipative phase shift in the density response (see below).

\section{Linear disc response for distant coplanar encounters}

\noindent Evaluating (\ref{psis}) at pericentre with the plane of the
secondary's orbit coplanar with the disc midplane we have

\begin{displaymath}
{\Psi}^{\prime} = -\frac{GM_{{\rm s}}}{q} \left[ 1 + \frac{1}{4}
{\left( \frac{r}{q} \right)}^2 (1-3\cos2\phi) \right] .
\end{displaymath}

\noindent The axisymmetric parts of this yield no net tidal effect
(Papaloizou \& Pringle 1977). The non--axisymmetric part
indicates a disturbance with azimuthal mode number $m=2$.  
Carrying out the potential expansion to terms of order $(r/q)^3$
yields terms describing modes with $m=1$ and $m=3$,
indicating that the dominant response comes from the $m=2$ component
with the $m=1$ and $m=3$ modes becoming increasingly significant in closer
encounters. This is in agreement with the findings of Korykansky \&
Papaloizou (1995), those
authors claim a satisfactory fit to their numerical results
can be calculated by considering only the Fourier components of the 
perturbative potential corresponding to modes $m=1$, $2$ and $3$;
the contribution from $m=1$ and $3$ modes being much smaller than that
from the $m=2$ mode.

\noindent Considering a potential perturbation

\begin{displaymath}
\Psi^{\prime}= \frac{3}{4} \frac{GM_{{\rm s}}}{q^{3}} r^{2} 
\exp{(2{\rm i}\phi)} ,
\end{displaymath}

\noindent the velocity and surface density perturbation amplitudes
are then given by:

\begin{displaymath}
V_{r{\rm a}}= 2{\rm i} \frac{GM_{{\rm s}}}{\Omega q^{3}} r ,
\end{displaymath}

\begin{displaymath}
V_{\phi {\rm a}}= - \frac{5}{4} \frac{GM_{{\rm s}}}{\Omega q^{3}} r ,
\end{displaymath}

\noindent and

\begin{displaymath}
\Sigma_{\rm a}= - \frac{(13-4n)}{4}
\frac{\mu \Sigma_{\rm 0}}{q^{3}} r^{3} .
\end{displaymath}

\noindent Here $\Sigma_{{\rm 0}}= \Sigma_{{\rm o}}
{(R_{{\rm o}}/r)}^{n-1}$ in which

\begin{displaymath}
\Sigma_{{\rm o}}= {\mathcal C}_{n} R_{{\rm o}}{\left[ 
\frac{\Omega_{{\rm o}}^{2}R_{{\rm o}}^{2}}{2K(n+1)} \right]}^{n}
{\left( \frac{H}{r} \right)}^{2n+1} .
\end{displaymath}

\noindent In the above; $\mu$ denotes the mass ratio $M_{{\rm
s}}/M_{{\rm p}}$,
the rotational velocity at the outer edge of the disc is
$\Omega_{{\rm o}}=\Omega(R_{{\rm o}})$ and we have considered $H/r$ to be a
constant (in practice $H/r$ is a weak power law in $r$, Larwood \&
Papaloizou 1997).
 
\noindent We can now deduce the part of the net torque acting on the
disc which gives rise to a change in the
magnitude of the disc angular momentum $J$. The rate of
change of the disc's angular momentum is then given
by (Papaloizou \& Pringle 1977):

\begin{displaymath}
\frac{{\rm d}J}{{\rm d}t}= -\int_{\rm disc} \Sigma^{\prime}
\frac{\upartial \Psi^{\prime}}{\upartial \phi}  r{\rm d}r{\rm d}\phi .
\end{displaymath}

\noindent Notice that the linear terms in the perturbed torque are
identically zero and that the net torque is second order in perturbed
quantities. Angular momentum transfer will then be a correspondingly
small effect in inviscid discs. In fact this part of the torque will itself
only be non-vanishing if the perturbation to the density becomes phase
shifted with respect to the perturbing potential such that
$\Sigma^{\prime} \propto \exp{m{\rm i}(\phi - \eta)}$.

\noindent In order to obtain an order of magnitude estimate of the
angular momentum exchange we consider the torque due to the leading
non-axisymmetric term of the potential expansion outlined above.
Integrating over the disc for the $m=2$ response, and from
$t= -\upi \Omega_{{\rm o}}^{-1}$ to $t=\upi \Omega_{{\rm o}}^{-1}$, we find

\begin{eqnarray}
\lefteqn{\uDelta J \equiv \frac{J-J_{\rm 0}}{J_{\rm 0}}} \nonumber\\
& & =-\frac{3\upi}{16} \frac{(13-4n)(7-2n)}{8-n}
\mu^{2} {\left( \frac{R_{{\rm o}}}{q} \right) }^{6} \sin 2\eta .
\label{dj1}
\end{eqnarray}

\noindent Note that we identify the angular momentum of the
unperturbed disc, $J_{{\rm 0}}$, with the angular momentum content of an
axisymmetric  polytrope:

\begin{displaymath}
J_{{\rm 0}}= 2\upi \int_{{\rm 0}}^{R_{{\rm o}}}
\Sigma_{{\rm 0}} r^{2} \Omega r{\rm d}r .
\end{displaymath}
  
\noindent In general a density perturbation is set up
in phase with the potential perturbation at a finite time before
pericentre passage. It is the differential rotation between the pattern
of the perturbation at the outside of the disc and the secondary
that results in a phase shift between the perturbing potential and the
density response. The existence of a phase lag allows a net angular
momentum transfer to occur between the secondary and the disc.

\noindent Our viscous numerical models also require consideration of the
angular momentum exchange due to viscous dissipation. This has
been derived by Papaloizou \& Pringle (1977):

\begin{equation}
\frac{{\rm d}J}{{\rm d}t}= -\int_{\rm disc} \frac{\Sigma_{{\rm 0}}
\Phi}{\Omega} 
r{\rm d}r{\rm d}\phi ,
\label{dj2}
\end{equation}

\noindent where
$\Phi \equiv 2 \nu ( e_{ij}e_{ij} - \frac{1}{3}e_{ii}^{2} )$
defines the rate of dissipation of mechanical energy per unit mass
in terms of the rate-of-strain tensor,
$e_{ij}$, of standard fluid theory. Evaluating (\ref{dj2}) in the
simple case of constant viscosity, and for the $m=2$ perturbations
calculated above, gives a contribution $[\uDelta J]_{\nu}$ to the
angular momentum exchange:

\begin{displaymath}
[\uDelta J]_{\nu} \approx
-70\upi \frac{7-2n}{15-2n} \left( \frac{\nu}{R_{{\rm o}}^{2}
\Omega_{{\rm o}}} \right)
\mu^{2} {\left( \frac{R_{{\rm o}}}{q}\right) }^{6} .
\end{displaymath}

\noindent Comparison of this result with equation (\ref{dj1}) implies
that for reasonable values of the polytropic index viscous torques
become important when $\eta \sim 10{\mathcal R}^{-1}$. Where
${\mathcal R}$ is the Reynolds' number
($\equiv r^{2}\Omega/\nu$) evaluated at the
outer edge of the disc. Since viscous timescales longer than
$\sim 10^{6} {\rm yr}$ are inferred in protostellar accretion
discs (Beckwith et al. 1990) we would require a dynamical phase
shift that is necessarily small compared with unity. There is no
reason why we should expect such a small value for $\eta$, hence
in general the viscous contribution to the process of angular momentum
exchange is expected to be negligible, as is confirmed by the
numerical results to be presented below.

\section{Extension to distant non--coplanar encounters}

\noindent Assuming the disc to be axisymmetric and the secondary-disc
interaction to occur only at pericentre, we need only consider the
inclination angle $\delta$ of the vector pointing to pericentre with respect
to the initial disc midplane. So if we choose the $x$-axis to lie
along the line of nodes, then ${\bmath D}$ is given by a simple rotation of
the previously considered coplanar trajectory about that axis:

\begin{displaymath}
{\bmath D}= q[4p,(1-4p^{2}) \cos \delta,(1-4p^{2})\sin \delta] .
\end{displaymath}

\noindent The perturbative potential
may then be recalculated and decomposed into a linear sum of
contributions corresponding to odd terms in $z$ (denoted by
${\psi}_{{\rm o}}$)
and even terms in $z$ (denoted by ${\psi}_{{\rm e}}$):

\begin{displaymath}
{\Psi}^{\prime}= {\psi}_{{\rm e}} + {\psi}_{{\rm o}} .
\end{displaymath}

\noindent Now, to terms of first order in $z$, equation (\ref{psis}) leads
us to consider non-axisymmetric components

\begin{displaymath}
{\psi}_{{\rm e}} = \frac{3}{4}\frac{GM_{{\rm s}}}{q^3} r^{2} {\cos}^{2}\delta
\exp(2{\rm i}\phi)
\end{displaymath}

\noindent and

\begin{displaymath}
{\psi}_{\rm o} = {\rm i}\frac{3}{2}\frac{GM_{\rm s}}{q^3}rz \sin2\delta
\exp({\rm i}\phi) .
\end{displaymath}

\noindent The non-axisymmetric part of the non--coplanar response that
is even in $z$, ${\psi}_{{\rm e}}$, carries an additional factor of
$\cos^{2}\delta$
in comparison with the analogous coplanar term. Hence
inclining the orbit diminishes the strength of the response found in
the coplanar case, being minimised in orthogonal configurations. Thus
the effectiveness of tidal truncation of the disc and the strength
of non--axisymmetric waves is reduced. In addition our estimate for
the angular momentum exchange is modified:
$\uDelta J \rightarrow \uDelta J \cos^{4} \delta$.

\noindent The part of the potential that is odd in $z$,
${\psi}_{{\rm o}}$, yields a disturbance with
azimuthal mode number $m=1$; this {\em bending mode} is found to warp
the disc and cause its
precession in circular orbit calculations (Papaloizou \& Terquem 1995,
Larwood et al. 1996). But
note that the amplitude we compute above is two times larger than that
for the analogous zero-frequency term in a circular orbit calculation
with $D=q$ (see Papaloizou \& Terquem 1995). Note also that we have
recovered a warping potential proportional to $\sin2\delta$, 
Ostriker (1994) showed that this result holds asymptotically for
distant encounters.

\subsection{Precession, warping and changes to the inclination}

\noindent We consider only the warp due to the part of the perturbative
potential with $m=1$, i.e. the longest wavelength mode. This warp will be
the longest--lived and most significant as long as the disc response is
linear. Also this component will cause precessional effects.

\noindent As before we calculate the forced response assuming the
perturbing potential to be fixed at its pericentre value.
We suppose that our coordinate system precesses at a rate
$\omega_{{\rm p}}$, which corresponds to the precession of the disc angular
momentum vector about the orbital
angular momentum vector of the secondary. In this case consideration
of the linearised fluid equations in their three-dimensional form gives the
response equation for the zero pattern speed forcing potential with
$m=1$ as (Papaloizou \& Lin 1995a) 

\begin{equation}
{{\rm d}\over {\rm d}r}
\left\{ {{\mathcal M}\Omega^2\over
\left[ \Omega^2 {(1-{\rm i}\epsilon)}^2-\kappa^2 \right] } 
{{\rm d}g\over {\rm d}r} \right\} =
{{\rm i}{\mathcal I} \over r},
\label{zerof}
\end{equation}
 
\noindent in which we have prescribed the complex function $g$,
defined through:

\begin{displaymath}
-{\rm i}rz\Omega^{2}g= \frac{P^{\prime}}{\rho} + \psi_{\rm o} .
\end{displaymath}

\noindent The $\epsilon$--term in (\ref{zerof}) is prescribed in order
to transform the location of the singularity due the resonant
denominator. This amounts to introducing a small phase shift which is
equivalent to $\alpha$ for a small shear viscosity, $\nu$. The remaining
symbols correspond to

\begin{equation} 
{\mathcal I}  = \int_{-H}^{+H}\frac{\rho
z}{c^2_{\rm s}}(\psi_{{\rm o}}+2{\rm i}\omega_{{\rm p}} 
rz\Omega \sin \delta ){\rm d}z ,
\label{force}
\end{equation}

\noindent and
 
\begin{displaymath}
{{\mathcal M}} = \int_{-H}^{+H} \rho z^2 {\rm d}z =
\frac{\Sigma {H}^{2}}{2n+3} .
\end{displaymath}

\noindent We employ the function $g$ to quantify the degree of warping manifest
in our numerical calculations. This is possible since a constant value of
$g$ corresponds to a rigid tilt of the disc, for which ${\bmath
r}\cdot{\bmath V}\equiv0$. Then for small amplitude warps
$g= V_{z}^{\prime}/V_{\phi}={\rm i}\zeta/r$,
where $\zeta(r)\exp({\rm i}\phi)$ is
the vertical displacement (Papaloizou \& Lin 1995a, Larwood \&
Papaloizou 1997). 

\noindent In the limit of slowly precessing small amplitude warps
we may neglect the contribution to (\ref{force}) from the
second term in parenthesis and make use of the identity

\begin{displaymath} 
\int_{-H}^{+H}\frac{\rho z^2}{c^2_{\rm s}} {\rm d}z
= \Sigma \Omega^{-2} ,
\end{displaymath}

\noindent derived by
integrating the equation of vertical hydrostatic equilibrium
(\ref{vert}). Then
to determine $g$ we need to integrate (\ref{zerof}) over the disc
twice. Carrying out this operation assuming $\kappa=\Omega$ and
$\alpha \gg \alpha^2$, we find the total range in the vertical
displacement $\uDelta(\zeta/r)$ to be

\begin{equation}
\uDelta(\zeta/r)= \frac{2n+3}{5-n}\alpha\mu
{\left(\frac{R_{\rm o}}{q}\right)}^{3}
{\left(\frac{r}{H}\right)}^{2}\sin2\delta .
\label{warp}
\end{equation}

\noindent If shear viscosity is a small effect then bending waves
propagate acoustically and should damp on the sound--crossing
timescale of the disc. This can provide a test of the importance
of any unquantified numerical viscous effects that might be
introduced due to the vertical shearing motions present in warps (see below).

\noindent We note also that in the above calculation there is an implicit
assumption that the sound-crossing time in the disc is small enough
that the warp is near its steady-state value after the encounter. This
is equivalent to the requirement that the interval over which the
interaction occurs is not less than the sound--crossing time for the
disc. If this condition is satisfied then it follows that the disc
will precess approximately as a rigid body (Papaloizou \& Terquem
1995) with a small frequency $\omega_{\rm p}$:

\begin{displaymath} 
\omega_{\rm p} \sin \delta =
\frac{\int \Sigma_{\rm 0}\frac{\upartial \psi_{\rm o}}{\upartial z}r^{2}
\sin\phi {\rm d}r{\rm d}\phi}{\int \Sigma_{\rm 0}r^3 \Omega 
{\rm d}r{\rm d}\phi} ,
\end{displaymath}

\noindent where the integrals are to be taken over the
area of the disc midplane. Evaluating from $r=0$ to $r=R_{\rm o}$ we find 

\begin{equation}
\frac{\omega_{\rm p}}{\Omega_{\rm o}}= -\frac{3}{4} \frac{7-2n}{5-n} \mu
{\left( \frac{R_{{\rm o}}}{q} \right)}^{3} \cos\delta .
\end{equation}

\noindent At large distances of pericentre global rigid body
precession is possible since the interaction interval
is $\sim 2\upi/\Omega_{\rm o}$.
Larwood et al. (1996) found for circular orbit binaries
that when the condition for rigid body precession was marginally
satisfied then the disc would precess differentially. This effect
amounts to a situation in which we have the differential precession of
sonically-connected annuli. For a parabolic encounter this would
result in a net precession of the disc with the main contribution
coming from the precession of an outer annulus of the disc.

\noindent In addition to the above, Papaloizou \& Terquem (1995) showed that
the $y$-component of the tidal torque $T_{{\rm y}}$ resulting
from secular terms
$\Psi^{\prime} \propto z$ gives rise to inclination evolution
according to

\begin{displaymath}
\frac{{\rm d}\delta}{{\rm d}t} \approx \frac{T_{y}}{J_{{\rm 0}}} .
\end{displaymath}

\noindent $T_{y}$ is identically zero unless we can introduce a phase
shift due to viscosity. Since
$\omega_{{\rm p}} \sim T_{x}/J_{{\rm 0}}$, we notice
that the relative importance of inclination evolution compared with
precessional effects in distant encounters is measured by the ratio

\begin{displaymath}
\frac{T_{y}}{T_{x}} \sim \alpha .
\end{displaymath}

\noindent Currently, values of $\alpha$ quoted for protostellar discs are
typically less than $\sim 10^{-2}$. Therefore we should
expect that as long as the disc response is linear then precessional
effects are more significant in determining the final orientation of
the disc than inclination changes.

\section{Numerical Simulations}

\noindent The numerical method we employ to study the dynamics
of viscous accretion discs is a version of SPH
(see Monaghan 1992, and references therein)
developed by Nelson \& Papaloizou (1993, 1994). 
The formulation uses a spatially variable {\em smoothing length},
associated with each particle, defined in
such a way as to ensure accurate energy conservation. Each
particle's smoothing length is defined to be half of the mean distance of the
six most distant nearest neighbouring particles chosen from a list
of forty--five members at each time--step.
 
\noindent In order to stabilize the calculations in the presence of shocks
an artificial viscous pressure term is included, according to the prescription
of Monaghan \& Gingold (1983). Although intended to prevent
particle penetration while giving positive definite dissipation, the practical
implimentation of the artificial viscosity
introduces a shear viscosity which results in disc spreading
much as in the standard theory of accretion
discs (Lynden-Bell \& Pringle 1974).

\noindent These standard accretion disc models were developed to study
tidal interaction and have been used in previously published
work (Larwood et al. 1996, Larwood \& Papaloizou 1997). Similar models
have been used by Artymowicz \& Lubow (1994).

\subsection{Basic numerical model}

\noindent The SPH disc models were constructed by generating random
position data for $17500$ particles, giving a zero-thickness disc of
unit radius centred on the origin of the computational coordinate
system (being coincident with the primary). Vertical positions were
generated in a similar fashion by equally seperating this initial
set of particles into seven uniformly-spaced planes of approximately constant
surface density, having total vertical extent $H_{{\rm o}}$. The
particles were then
allocated velocities according to a softened Keplerian potential:
$\Omega= (r^{2} + b^{2})^{-3/4}$, the softening length $b=0.2$ chosen
to avoid the singularity of a Keplerian potential that occurs near the
origin. It is then only required to determine the polytropic constant,
$K$, from the equation
of vertical hydrostatic equilibrium. We choose $K$ to give a total vertical
semi-thickness $H_{{\rm o}}$ at the initial outer radius,
given that we employ a polytrope of index $n=1.5$ in our numerical
models (which ensures $\nu \approx {\rm constant}$, Larwood et al. 1996).

\noindent Units are such that $G=M_{\rm p}=1$, giving a time unit
of $\Omega_{{\rm o}}^{-1}$. The affect of numerical viscosity is to produce an
effective shear viscosity which we calibrate against a standard constant
viscosity model. The results of our calibration exercise imply

\begin{displaymath}
\alpha \sim 0.008 \times
{\left( \frac{R_{{\rm o}}}{r} \right)}^{1/2}
{\left( \frac{H}{r} \right)}^{-2/3} .
\end{displaymath}

\noindent Before introducing the secondary the disc models were
allowed to relax for a few complete rotations solely in the gravitational
field of the primary and under SPH pressure and viscous forces. This
is required to establish an
approximate equilibrium state of the system. Thick disc models with
$H/r= 0.15$ (relevant to
protostellar discs) as well as thin disc models with $H/r= 0.05$ (chosen
to test discs with a significantly smaller sound speed) are
considered. Due to the different thicknesses and the nature of the SPH
numerical viscosity the thick
and thin disc models also represent different viscosity regimes
(Larwood \& Papaloizou 1997), with
$\nu \sim 10^{-3}$ in the thick disc case and
$\nu \sim 10^{-4}$ for the thin disc models.
As a result the outer
radius of the disc relaxes to different values in each case. For the
thin disc models we take $R_{{\rm o}}= 1.2$ and for the thick disc models
we use $R_{{\rm o}}= 1.4$.

\noindent The secondary is introduced at a distance of $10$ units from
the primary, corresponding to a time $\sim -15$, in
prograde rotation with the disc. The secondary's position is
given from the computed time at each SPH integration step
by the expressions derived in Section $2.2$. The potential due
to the secondary is then calculated from equation (\ref{secpot}).
These expressions for the secondary are used in their softened form.

\noindent Parameters considered are for secondaries with unit mass
ratio ($M_{{\rm s}}=1$) and distances of closest approach $q= 2.4$, $3.6$,
$4.8$ and $6.0$. Both coplanar and non-coplanar cases with $\delta= 15\degr$,
$30\degr$ and $45\degr$ are tested.

\subsection{Calculation of the disc surface density}

\noindent We obtain an azimuthally averaged disc surface density
profile by considering the particle distribution over $140$ concentric
circular annuli of width $0.1$, taken from radius $r= 0.1$ to $1.5$
in the midplane of the disc. The surface density at the location of
the centre of each radial bin
is ascribed the mean value determined by projecting particles onto
the midplane, summing the number in each annulus and
dividing this number by the area of the annulus. Surface densities
thus obtained are expressed in units of the disc's mean surface density
at initialisation.

\subsection{Determining changes to the disc orientation}

\noindent As in previous work, we introduce angles $\iota$ and $\Pi$
to quantify respectively the effects of warping and of precession in
our numerical data.

\begin{displaymath}
\cos \iota = \frac{{\bmath J}_{{\rm A}} \cdot {\bmath J}_{{\rm D}}}
              {| {\bmath J}_{{\rm A}} ||{\bmath J}_{{\rm D}}|}
\end{displaymath}

\noindent and

\begin{displaymath}
\cos \Pi = \frac{({\bmath J}_{{\rm O}} {\bmath \times}
            {\bmath J}_{{\rm D}}) \cdot {\bmath u} }
            {| {\bmath J}_{{\rm O}} {\bmath \times}
            {\bmath J}_{{\rm D}}| |{\bmath u}|} .
\end{displaymath}

\noindent ${\bmath J}_{{\rm D}}$ is the disc angular momentum calculated as the
sum over all disc particle angular momenta, ${\bmath J}_{{\rm A}}$ is
the angular
momentum within an annulus. The arbitrary reference vector
${\bmath u}$ lies in the
orbital plane of the secondary. For convenience we choose ${\bmath u}$
such that $\Pi$ takes the initial value of $\upi / 2$ radians, its
subsequent evolution directly measuring the disc precession.

\noindent Retrograde
precession of ${\bmath J}_{{\rm D}}$ about ${\bmath J}_{{\rm O}}$
implies that the angle $\Pi$
decreases linearly with time, descirbing an arc in the orbital plane of the
secondary. Applying the pericentre precession frequency for one
outer--disc
orbital period, we estimate the small change in the disc precession
angle expected to occur in a $n=1.5$ polytrope as $\uDelta \Pi$:

\begin{displaymath}
\uDelta \Pi = -\frac{12\upi}{7} \mu
{\left( \frac{R_{\rm o}}{q} \right)}^{3} \cos\delta .
\end{displaymath}

\noindent During the parabolic fly--by of a secondary body, the linear
response of a fluid disc is to precess approximately as a rigid body
(see below). The small amount
of precession results in a shift of the disc midplane away
from its initial configuration, providing an apparent relative tilt
with respect to the initial midplane, whilst
maintaining the angle between the disc and orbital angular momentum vectors.
For a small change in the precession angle $\uDelta \Pi$, producing a small
tilt $\tau$, we expect

\begin{displaymath} 
\tau = |\uDelta \Pi| \sin \delta .
\end{displaymath}

\noindent Typical model parameters are $\mu=1$, $q=4R_{{\rm o}}$ and
$\delta=30\degr$, giving a tilt $\sim 2\degr$,
being much larger than the expected change to the inclination $\delta$.

\noindent We note that for small values, the total range in $\iota$ (being the
angle between the angular momentum vector of the disc material contained within
a specified cylindrical annulus and the total angular momentum vector of the
disc) is equivalent to $\uDelta(\zeta / r)$.

\section{Numerical Results}

\noindent All our disc models lost angular momentum to the secondary,
resulting in a negative sign in the net angular momentum exchanges. Our
thick disc models, having the larger outer radius,
showed the largest loss of angular momentum; our thin disc models,
having a smaller outer radius, lost less angular
momentum. Non-axisymmetric waves with azimuthal mode number $m=2$
were visibly excited in all but the most distant encounters,
as we would expect given that the $m=2$ inner Lindblad resonance with pattern
speed $2\omega_{{\rm o}}$ lies far from the disc edge when $R_{{\rm o}}/q$ is
sufficiently small. 

\noindent Additionally, in non-coplanar configurations, we found that
the disc became warped and showed precessional effects and inclination changes.
The affect on the response of introducing a finite orbital inclination was
generally to diminish the effectiveness of the secondary's tide,
although there were some indications that high inclination encounters could
enhance angular momentum transfer.

\subsection{Coplanar encounters}

\subsubsection{Angular momentum exchange}

\noindent The parameters for these models are given in Table
(\ref{table1}), along with the corresponding net angular momentum
changes found in the disc models.
Figure (\ref{fig1}) shows a particle projection plot for the
disc particles in model $3$, taken at a time for which the secondary is
close to pericentre. The angular separation between the longitude
of the secondary and the major axis of the disc's elliptical
envelope is $\sim \upi/4$ radians. If we identify this with the
dynamical phase shift, $\eta$, then we would have a much larger phase shift
than would be expected from considering dissipative effects alone.
Hence the disc
viscosity would not be expected to make a significant contribution
to the process of angular momentum exchange in these models. In fact
our results indicate that the difference in the magnitudes of the angular
momentum exchanges between the thick and the thin disc models
can be explained solely in terms of the different radial
extents associated with the two types of model.


\begin{table}
\centering
\begin{minipage}{48mm}
\caption{The change in the angular momentum content of the disc
for model parameters used in coplanar encounters.}
\begin{tabular} {ccccccccc} \hline \hline

model & $H/r$ & $q$ & $\uDelta J$ \\ 
\hline

 1 & 0.15 & 2.4 & -1.6$\times 10^{-1}$ \\
 2 & 0.15 & 3.6 & -3.0$\times 10^{-2}$ \\
 3 & 0.15 & 4.8 & -2.5$\times 10^{-3}$ \\
 4 & 0.15 & 6.0 & -7.6$\times 10^{-4}$ \\
\hline
 5 & 0.05 & 2.4 & -1.4$\times 10^{-1}$ \\
 6 & 0.05 & 3.6 & -1.1$\times 10^{-2}$ \\
 7 & 0.05 & 4.8 & -3.8$\times 10^{-4}$ \\
 8 & 0.05 & 6.0 & -1.9$\times 10^{-5}$ \\

\hline \hline
\end{tabular}
\label{table1}
\end{minipage}
\end{table}


\noindent The magnitude of $\eta$ at pericentre is determined by
the angular velocity of the secondary in the frame rotating with the
outer disc, the former varying in time. As a result models which
have smaller (larger) $q$-values
show smaller (larger) $\eta$-values. For the range of
$q$-values considered here we shall take the approximate median value
$\eta=\upi/4$ in evaluating our analytically derived expressions,
noting that this gives maximal angular momentum transfer.

\begin{figure}
\vspace{70mm}
\caption{Particle positions projected onto the $(x,y)$ plane for model
$3$, given at a time for which the secondary is close to pericentre. Each
point represents a particle position, the secondary's position is 
denoted by an open circle and its projected path by a dotted line.}
\label{fig1}
\end{figure}

\noindent We present plots of the total angular momentum of the disc
versus time for models $5$ to $8$ in Figure (\ref{fig2}).
Models $5$ and $6$ show an impulsive form for the angular momentum
exchange which takes place over $\sim 2\upi \Omega_{{\rm o}}^{-1}$ time
units about
pericentre passage. This simple picture shows signs of breakdown
in models $7$ and $8$, which indicate in addition a
significant {\em return} of angular momentum from the secondary to
the disc occuring after the initial interaction described above. This reversal
in the direction of the angular momentum transfer can be
understood in terms of the passage of the outer disc through the point
where the $m=2$ elliptical pattern and the longitude of the secondary
are $90\degr$ out of phase (giving zero net torque
at that instant). We can also say that we expect the return of angular
momentum to be more effective for parabolae with larger $q$-values,
since in those cases $R_{{\rm o}}/D$ varies more slowly in time. Indeed, as
the distance of closest approach becomes sufficiently large,
the net angular momentum exchange should tend to zero as the disc
responds {\em adiabatically} to the perturbation. In the case of a
viscous disc the angular momentum exchange would tend to the small but
finite value expected from viscous dissipation alone.

\noindent Our analytical expression for $\uDelta J$ does not include
the affect of the return of angular momentum to the disc. We
compare values derived from equation (\ref{dj1}) with the angular
momentum exchanges we determine in our numerical models, considering
only the part of the interaction in which the disc initially loses angular
momentum (see Figure \ref{fig3}). We note that an additional
return of angular momentum to the disc will modify the magnitude of the
resultant exchange by a factor which depends on $q$.

\noindent As we should expect, our analytically derived expression
gives the best agreement with our numerical results for the most
distant encounters. However the expression slightly over-estimates these
values, which is probably a consequence of choosing a phase shift that
gives maximal angular momentum transfer. The numerical values for $\uDelta J$ 
are under-estimated by the predicted values as the encounters become closer.
This probably owes less to the exclusion of $m=1$ and $m=3$
modes from our calculations than to the onset of
non-linearity in the disc response, which in some cases has been
found to double the angular
momentum transfer seen in similar but purely linear calculations
(Korykansky \& Papaloizou 1995).

\begin{figure*}
\vspace*{120mm}
\caption{The angular momentum of the disc, in units of its initial
value, versus time for models $5$--$8$.}
\label{fig2}
\end{figure*} 

\begin{figure}
\vspace{60mm}
\caption{The magnitude of the angular momentum initially lost by the
disc for models $1$--$8$. The thick disc data is denoted by open
squares, the thin disc data by open triangles and the analytical
result by a solid line.}
\label{fig3}
\end{figure} 

\begin{figure}
\vspace{60mm}
\caption{Surface density profiles for model $2$, denoted by solid
circles, and an unperturbed model, denoted by open circles. Each is
taken for a total run time of about $30$ units.}
\label{fig4}
\end{figure}

\subsubsection{The redistribution of disc matter}
 
\noindent In all cases with $q \ge 4R_{{\rm o}}$, the surface density
profile at equal times (examined at a total time elapsed of about $30$ units)
was found to be barely distinguishable from that of an unperturbed model.
For $q \sim 3R_{{\rm o}}$ there were indications that tidal truncation of the
disc had occured in much the same way as would
be expected for a system involving a bound secondary (Lin \& Papaloizou
1979a). This case carries a weakly non--linear component to the
response with the negative angular momentum flux into the disc
being sufficiently large so as to result in local non-linear
dissipation near the disc edge. This
behaviour of the system confirms that the break-down of linearity
occurs for $q < 4R_{{\rm o}}$ (cf. Hall et al. 1996). 
The affect of tidal truncation on the disc surface density profile
is represented in Figure~(\ref{fig4}). The middle part of the
perturbed disc
was sculpted into a more compact and uniformly distributed structure
and the outer disc was abruptly cut-off. This feature will only be
subject to significant subsequent evolution on the viscous timescale.

\begin{figure*}
\vspace*{80mm}
\caption{Positional $(x,y)$ data for model $1$ taken at a time $t=+13$.
The same data is displayed at two different scales.}
\label{fig5}
\end{figure*} 

\begin{figure*}
\vspace*{120mm}
\caption{Positional $(x,z)$ and $(y,z)$ data for model $1$ taken at a
time $t=+13$. Only particles with $r < 1.2$ are plotted.}
\label{extra1}
\end{figure*} 

\noindent In models $1$ and $5$
with $q \sim 2R_{{\rm o}}$ the disc response was strongly non-linear, the
surface density structure being severely disrupted and
truncated. As indicated in Figure~(\ref{fig5}) a material arm of
several disc radii in extent is ejected from the outer-edge of the
disc and the secondary appears to capture some of the disc
particles. Much of the remaining disc material is re-formed into a dense ring.
The formation of a ring in both of these models is accompanied by a
suppression of central mass accretion. This indicates that the ring
forms by the
reflection of an inwardly propagating trailing wave, carrying a
negative angular momentum flux, producing an outwardly propagating
leading wave carrying negative angular momentum
from the inner region of the disc. The reflected wave then encounters
the rear portion of its inwardly propagating
counterpart and dissipation occurs. The net effect is to
clear an inner annulus of material, form a ring and suppress central
mass accretion.

\noindent The production of circumprimary rings in close
binary encounters has been observed before in numerical simulations
(Clarke \& Pringle 1993). It is understandable that the formation of a
ring is only found in very strong encounters
since it is in these cases that the $m=2$ spiral density waves,
launched at
the outer disc, are strong enough to propagate to sufficiently small
radii that their radial wavelength falls
below the resolution limit of any code that is used to model the disc
matter. We find that the
resolution at the estimated radius of reflection is marginal given the
predicted radial wavelength (given by equation \ref{disp2}); hence we
conclude that ring formation is observed in these simulations
but that it is possibly explicable as a numerical effect.

\noindent Accompanying these phenomena we also find that the envelope
of the disc takes on an elliptical shape, however unlike the $m=2$ bar
that is generated prior to strong interaction, this form is not
centred on the primary. This feature does not disperse
on a dynamical timescale which implies that the disc has become
genuinely elliptical. The implication is that the
disc has taken on an asymmetric structure that will only disperse on
the viscous timescale. In Figure (\ref{extra1}) we plot the side view
particle projections corresponding to Figure (\ref{fig5}). 

\noindent In models $1$ and $5$ the
secondary removed $\sim 5 \%$ of the initial disc mass (i.e. about
$10^3$ particles), most of which was {\em captured}. The captured
material seemed to consistently manifest as a dense
clump of eccentrically orbiting material (see Figure
\ref{fig5}). Further investigation of this phenomenon shall be
the subject of future studies. In model $2$ with
$q \sim 2.6R_{{\rm o}}$ only $\sim 10$
particles were lost. In models $3$, $4$, $7$ and $8$, for which
$q \ge 3R_{{\rm o}}$, no material was lost from the disc.

\subsection{Non--coplanar models}

\subsubsection{Angular momentum exchange}

\begin{figure}
\vspace{60mm}
\caption{The angular momentum of the disc, in units of its initial
value, versus time for models $3$, $3a$, $3b$ and $3c$; represented
by squares, crosses, diamonds and pluses respectively.}
\label{fig6}
\end{figure} 

\begin{figure}
\vspace{60mm}
\caption{The angular momentum of the disc, in units of its initial
value, versus time for models $8$ and $8c$; represented by squares
and crosses respectively.}
\label{fig7}
\end{figure} 

\begin{figure}
\vspace{60mm}
\caption{The magnitude of the angular momentum initially lost by the
disc for non-coplanar models. The thick disc data
is denoted by squares, the thin disc data by
triangles and the analytical result by a solid line.}
\label{fig8}
\end{figure}

\noindent The data for our non-coplanar models is given in Table
(\ref{table2}). In Figure (\ref{fig6}) we show the angular momentum
evolution of the disc for models $3$, $3a$, $3b$ and $3c$. As the
orbital inclination angle is increased the magnitude of the initial
transfer of angular momentum from the disc to the secondary decreases.
Hence inclining the orbit of the secondary results in a weaker interaction.
In addition the magnitude of the angular momentum returned from the
secondary to the disc, compared with the net exchange in each case,
is also reduced. This is because the mean distance of the secondary
from the disc edge is larger at higher inclinations.
We illustrate an extreme in this behaviour by examining the 
angular momentum evolution for models $8$ and $8c$ in
Figure (\ref{fig7}). Although the initial
transfer is largest in the coplanar model the {\em net} transfer is
larger for the non-coplanar model.

\noindent As for the coplanar case we compare the inferred $\uDelta J$ values,
determined for the initial phase of the angular momentum exchange,
with our expression derived from linear theory and modified
for inclined orbits of the type considered here. We present these
results in Figure (\ref{fig8}). The general behaviour of
the numerical values compared with the analytical result is similar to that
found in the coplanar case. In addition it appears that low
inclination models give better agreement with the analysis than
the high inclination models. This is probably due to the enhancement of the
angular momentum flux through the non-coplanar $m=1$ contribution
to the response, which is most important at high inclinations
(Papaloizou \& Terquem 1995). We note also that the thin disc models
appear to show larger values than the thick disc cases, this is
probably due to a greater tendency to non-linearity in the response
which could be more significant in the presence of the vertical shearing
motions found in warps.

\begin{figure}
\vspace{60mm}
\caption{Surface density profiles for model $2$, denoted by solid
circles, and model $2c$, denoted by open circles. Each is
taken for a total run time of about $30$ units.}
\label{fig9}
\end{figure} 

\begin{figure}
\vspace{60mm}
\caption{Precession angle data versus time for model $3c$.}
\label{fig11}
\end{figure} 


\begin{table*}
\centering
\begin{minipage}{106mm}
\caption{Data for model parameters used in non-coplanar
encounters. $\uDelta \delta$ denotes the change to the orbital
inclination of the disc, all other symbols are defined in the text. Angles
are expressed in radians except for $\delta$ and the quantities in
parentheses which are expressed in degrees.}
\begin{tabular} {cccccccc} \hline \hline

model & $H/r$ & $q$ & $\delta$ & $\uDelta J$ & $\uDelta \delta$ & $\uDelta \Pi$ & $\tau$ \\ 
\hline
 1a & 0.15 & 2.4 & 15 & -1.5$\times 10^{-1}$ & 1.7$\times 10^{-2}$ &
-0.19 & 4.9$\times 10^{-2}$ (2.8) \\
 1b & 0.15 & 2.4 & 30 & -1.3$\times 10^{-1}$ & 2.8$\times 10^{-2}$ &
-0.17 & 8.5$\times 10^{-2}$ (4.9) \\
 1c & 0.15 & 2.4 & 45 & -9.7$\times 10^{-2}$ & 2.7$\times 10^{-2}$ &
-0.15 & 1.1$\times 10^{-1}$ (6.1) \\
 2a & 0.15 & 3.6 & 15 & -2.0$\times 10^{-2}$ & 4.4$\times 10^{-3}$ &
-0.11 & 2.8$\times 10^{-2}$ (1.6) \\
 2b & 0.15 & 3.6 & 30 & -1.9$\times 10^{-2}$ & 6.1$\times 10^{-3}$ &
-0.10 & 5.0$\times 10^{-2}$ (2.9) \\
 2c & 0.15 & 3.6 & 45 & -8.7$\times 10^{-3}$ & 5.0$\times 10^{-3}$ &
-0.08 & 5.7$\times 10^{-2}$ (3.2) \\
 3a & 0.15 & 4.8 & 15 & -2.1$\times 10^{-3}$ & 3.6$\times 10^{-4}$ &
-0.08 & 2.1$\times 10^{-2}$ (1.2) \\
 3b & 0.15 & 4.8 & 30 & -1.6$\times 10^{-3}$ & 4.9$\times 10^{-4}$ &
-0.07 & 3.5$\times 10^{-2}$ (2.0) \\
 3c & 0.15 & 4.8 & 45 & -1.1$\times 10^{-3}$ & 2.7$\times 10^{-4}$ &
-0.05 & 3.5$\times 10^{-2}$ (2.0) \\
 4a & 0.15 & 6.0 & 15 & -2.0$\times 10^{-4}$ & 0.6$\times 10^{-4}$ &
-0.06 & 1.6$\times 10^{-2}$ (0.9) \\
 4b & 0.15 & 6.0 & 30 & -1.9$\times 10^{-4}$ & 0.7$\times 10^{-4}$ &
-0.05 & 2.5$\times 10^{-2}$ (1.4) \\
 4c & 0.15 & 6.0 & 45 & -1.5$\times 10^{-4}$ & 0.3$\times 10^{-4}$ &
-0.04 & 2.8$\times 10^{-2}$ (1.6) \\
\hline
 5c & 0.05 & 2.4 & 45 & -7.2$\times 10^{-2}$ & 2.7$\times 10^{-2}$ &
-0.15 & 1.1$\times 10^{-1}$ (6.1) \\
 6c & 0.05 & 3.6 & 45 & -4.1$\times 10^{-3}$ & 1.4$\times 10^{-3}$ &
-0.07 & 4.9$\times 10^{-2}$ (2.8) \\
 7c & 0.05 & 4.8 & 45 & -2.4$\times 10^{-4}$ & 0.2$\times 10^{-3}$ &
-0.05 & 3.5$\times 10^{-2}$ (2.0) \\
 8c & 0.05 & 6.0 & 45 & -5.3$\times 10^{-5}$ & 0.3$\times 10^{-3}$ &
-0.04 & 2.8$\times 10^{-2}$ (1.6) \\

\hline \hline
\end{tabular}
\label{table2}
\end{minipage}
\end{table*}


\begin{figure*}
\vspace*{80mm}
\caption{Positional data for model $1c$ at a time $t=+13$.}
\label{fig10}
\end{figure*} 

\subsubsection{The redistribution of disc matter}

\noindent Tidal truncation of the model discs was found to operate
for non-coplanar models as for the coplanar cases, but
occurring at a reduced level for larger orbital inclinations.
In Figure (\ref{fig9}) we compare the surface density profiles for
models $2$ and $2c$, examined at a total time elapsed of $30$ units
from the time of the introduction of the secondary (cf.
Figure \ref{fig4}). The surface density profile for
model $2c$ appears to be intermediate between the coplanar case and
the unperturbed model with the outer disc not as strongly cut-off
as in the coplanar case but more so than in the unperturbed case
(which is not cut-off at all).   

\noindent Mass transfer streams and spiral arms were generated as
in the coplanar models except that they could become warped and even
{\em wrapped} (cf. Clarke \& Pringle 1993) about the disc, for the
closest encounters, as shown in Figure (\ref{fig10}).
Mass loss from the disc was also of a similar order to that observed
in the coplanar models.    

\subsubsection{Precession, warping and inclination changes}

\noindent All our non-coplanar models showed a small amount of
precession. In Figure (\ref{fig11}) we give the precession angle
evolution for model $3c$. As for the torque which gives rise to
angular momentum exchange the torque giving precession of the disc
is found to assert itself essentially as an impulse at pericentre.
The small amount of precession manifest in all these models
was found to give a small relative tilt $\tau \sim 1\degr-6\degr$,
even at the largest $q$-value, in reasonable agreement with the
expression derived above. We give particle projection plots
of model $6c$ in Figure (\ref{fig12}). It is clear that the midplane
of the disc has become shifted with respect to its initial orientation.

\noindent As well as showing precessional behaviour the model discs
also became warped. We compare the size of the warps found in each
model at a time for which the companion has receeded to $r=10$,
with the equation (\ref{warp}), in Figure (\ref{fig13}). The expression
derived from linear
theory seems reliable for distant encounters in the thick disc
models. In the thick disc cases with $q<4R_{{\rm o}}$ the
magnitude of the warp is over-estimated,
consistent with non-linear damping of strong vertical shearing
motions.

\begin{figure*}
\vspace*{120mm}
\caption{Positional data for model $6c$, the top frame gives the
unperturbed particle configuration and the bottom frame gives the
positions at a time $t=+28$ after pericentre passage.}
\label{fig12}
\end{figure*}

\noindent The size of the warps present in the thin disc models seem to obey
a proportionality relation with the tidal strength of the encounter
but are over-estimated by a factor of $\sim 10$. The three times
larger sound-crossing time
($\equiv \int \bar{c}_{{\rm s}}^{-1} dr$) in the thin disc models
prevents saturation of the warps on the encounter timescale.
In Figure (\ref{fig14}) we give the warp evolution for
model $3c$. A decay timescale of approximately $60$ units is inferred,
being about
$5$ sound-crossing times. A similar
result was found to hold for the thin disc models. This is consistent
with bending modes that propagate and damp as acoustic waves
(Papaloizou \& Lin 1995a).

\noindent Inclination changes to the disc were typically small in cases such
that $q > 3R_{{\rm o}}$, for $q \sim 3R_{{\rm o}}$ and
$q \sim 2R_{{\rm o}}$ the
inclination changes were more significant at $\sim 1 \%$ and
$\sim 10 \%$ respectively. The net change to the inclination was found
to be much smaller than the tilt due to precession in all cases.

\begin{figure}
\vspace{60mm}
\caption{The magnitude of the range in the vertical elevation at
a time of about $30$ units. The thick disc data for models with
$q\ge4R_{\rm o}$ is denoted by open
squares, thick disc data for models with
$q\sim3R_{\rm o}$ by open circles, the thin disc data by
asterisks and the analytical result for the thick disc models
by a solid line.}
\label{fig13}
\end{figure} 

\begin{figure}
\vspace{60mm}
\caption{The warp evolution shown as the inclination versus radius
for approximately equal time intervals. Squares represent data at time
$t=+31$, pluses represent data at time $t=+39$, triangles represent data
at time $t=+50$, crosses represent data at time $t=+61$ and diamonds
represent data at time $t=+70$.}
\label{fig14}
\end{figure}

\section{Discussion}

\subsection{Summary}

\noindent In this paper we have presented the results of
analytical calculations and
hydrodynamic simulations of the tidal interaction of a point
mass encountering a viscous accretion disc on a parabolic prograde fly-by.
We have considered the interaction for the case when the orbital plane
of the secondary coincides with the midplane of the disc as well
as the case when the two planes are inclined.

\noindent We have shown that for coplanar systems the most
important component of the
perturbing potential that gives a net tidal effect is the $m=2$
bar mode applied at pericentre. Further to this we have demonstrated
that the affect of introducing an anomolous shear viscosity of the magnitude
thought to be relevant to protostellar accretion discs is only significant
for very distant encounters.
The exchange of angular momentum occuring between the secondary
and the accretion disc is calculated from simplified inviscid
considerations by modelling the interaction as a non-resonant impulse
occurring at pericentre. A net torque is found to act on the disc due to
the generation of a phase lag in the density response. To calculate
the disc response we apply the pericentre torque over a finite
interaction interval for a fixed value of the phase shift.

\noindent In our non-coplanar models we have shown in addition
that a $m=1$ bending mode is excited whose action is to warp the
disc, cause it to precess away from its original plane and give
a small change to the inclination. The precession of the
disc produces a small tilt of the disc midplane relative to its
initial configuration, this effect is always more important than the
inclination change in determining the final orientation of the disc.
The size of the tilt is significant even for the weakest
encounters considered, namely those with the distance of closest
approach of order five times the outer radius of the disc, for which
the exchange of angular momentum is a very small fraction of the total
disc content. There is also evidence to suggest that the $m=1$
component of the perturbing potential can become as important as the
$m=2$ component in the angular momentum exchange at high inclinations.

\subsection{Consequences for protostellar accretion discs}

\subsubsection{Silhouette discs in the Orion Nebula}

\noindent Clarke \& Pringle (1991) estimated the encounter rate,
$\Gamma$, for parabolic point masses encountering discs. For the
densest parts of the Trapezium cluster they estimated $\Gamma \sim
0.1 {\rm Myr^{-1}}$ for discs with $R_{{\rm o}} \sim 100 {\rm AU}$.
They argued that
since disc lifetimes are $\sim 1 {\rm Myr}$ then disc penetrating encounters
are rare. We note that approximately $\Gamma \propto q^2$. The
implication of this is that encounters with $q \sim 3R_{{\rm o}}$ are the
rule rather than the exception.

\noindent In our numerical models encounters with $q \sim 3R_{{\rm o}}$
produce disc tilts $\sim 3\degr$ and show tidal truncation such
that the discs become more compact with strongly cut-off edges.
Directly imaged silhouette discs in the Trapezium cluster are found
to exhibit a similar property in their surface density profiles
(McCaughrean \& O'Dell 1996). Also the largest
silhouette disc ({\em Orion 114-426}; with
$R_{{\rm o}}\sim 500{\rm AU}$) which is
almost edge-on to the line of
sight, shows a clear radial asymmetry and vertical distortion. But
note that
other environmental influences cannot be ruled out as explainations
for these features of the observations (O'Dell, Wen \& Hu 1993,
McCaughrean \& O'Dell 1996).

\subsubsection{Discs in eccentric binary systems}

\noindent Korykansky \& Papaloizou (1995) noted that it is possible
to model the accretion disc response in an eccentric binary system
with a succession of parabolic pericentre passages. In this picture
the accretion disc suffers a series of impacts in which it loses
angular momentum to the secondary, thus reducing its lifetime. The
secondary would increase its eccentricity with every impact.

\noindent In coordinates based on the primary, a secondary with specific
angular momentum $L$ and orbital eccentricity $e$ has a pericentre distance
$q$:

\begin{displaymath}
q= \frac{1}{1+e} \frac{L^2}{G(M_{{\rm p}}+M_{{\rm s}})} .
\end{displaymath}    

\noindent Therefore at pericentre we can write the change in the
eccentricity $\udelta e$ in terms of the change to the specific
angular momentum content of the disc $\udelta J$:

\begin{displaymath}
\udelta e= -2(1+e) \frac{\udelta J}{L} .
\end{displaymath}    

\noindent If we then consider the simplified case with the mass of the
disc distributed in a thin ring of radius $R_{{\rm o}}$, being in Keplerian rotation
about the primary alone, then we find:

\begin{displaymath}
\udelta e= -2\uDelta J \left(\frac{R_{{\rm o}}}{q}\right)^{1/2}
\sqrt{\frac{1+e}{1+\mu}} .
\end{displaymath}    

\noindent Then considering further extreme parameters for angular
momentum exchange, such as $\mu \sim 1$, $e \sim 1$
and $q \sim 2R_{{\rm o}}$ we find that $\udelta e \sim \uDelta J$.
Thus if we take the values for $\uDelta J$ present
in our models we would expect that eccentricity changes to the
secondary are generally less significant. The
implication of this is that an eccentric binary may remove a
significant fraction of the disc's angular momentum without
becoming unbound.

\subsection{Consequences for planetary systems}

\subsubsection{The Solar system}

\noindent If planets form near the midplane of an accretion disc and
if protoplanetary discs can become tilted by some mechanism without
exerting a significant torque upon the central star, then it is
reasonable to suppose that planetary systems are able to form with
non-zero inclinations to the stellar equator. We note that it is
trivial to show from standard astronomical data that the
invariable plane of the Solar system is tilted by
$\sim 6\degr$ with respect to the Solar equator.

\noindent The largest tilt angles we found in our simulations were
$\sim 6\degr$ for $q \sim 2R_{{\rm o}}$. It is therefore plausible that
a unique close encounter in the early history of the Solar system
could have caused the Solar obliquity.
If on the other hand we were to suppose a number of more distant but
coherent encounters then we must appeal to the existence of an eccentric
Solar binary companion which has long since become unbound through its
tidal interaction with the primitive Solar disc. However, our result of
the previous Subsection indicates that unless it had an eccentricity
very close to unity, a Solar binary should still be in evidence.

\subsubsection{The Beta Pictoris system}

\noindent A gaseous disc in which there already exists
a planet located within a tidally cleared gap (Lin \& Papaloizou 1979b)
interacts with that planet through gravitational torques.
If the outer disc becomes tilted due to a close binary encounter then
the planetary orbit would not necessarily tilt rigidly with the rest
of the disc. A relative inclination between the disc and the planetary orbit
would be generated. We note that the dusty circumstellar disc
associated with Beta Pictoris shows an inner warp
$\sim 3\degr$ (Burrows, Krist \& Stapelfeldt 1995). This has been
modelled with an inclined planet orbiting interior to the
disc (Mouillet et al. 1997). Asymmetries on the scale of the disc
have also been noted and a stellar encounter proposed as a possible
mechanism for their generation (Kalas \& Jewitt 1995).

\section*{Acknowledgments}

This work was supported by PPARC grant GR/H/09454 and the author is
supported by a PPARC studentship. Thanks also go to Dr. Mark McCaughrean for
useful discussions on the silhouette disc observations and to
Professor John Papaloizou for his support and advice throughout
this project. Dr. Caroline Terquem is thanked for providing
instructive and insightful comments as the referee of this paper.
 


\label{lastpage}
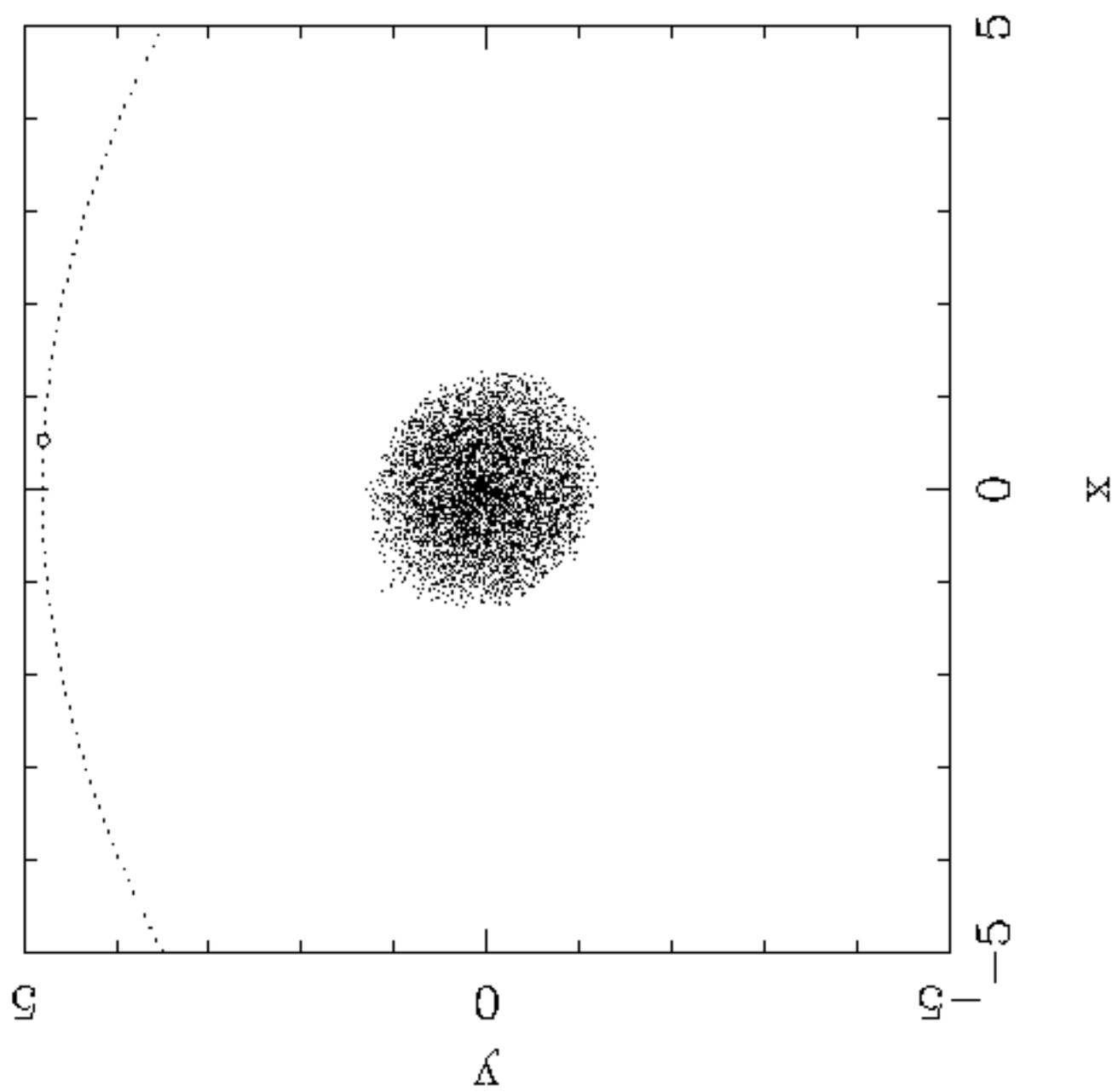
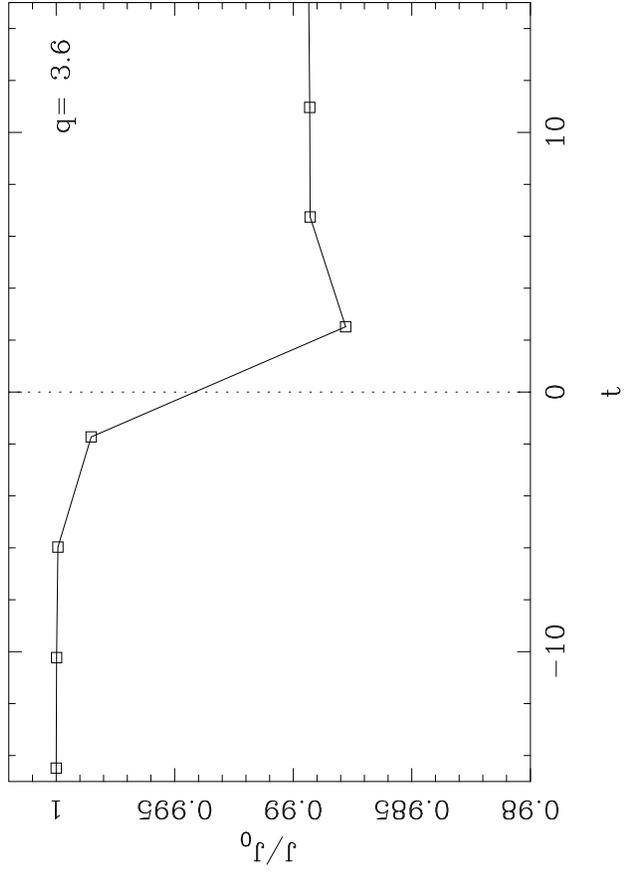
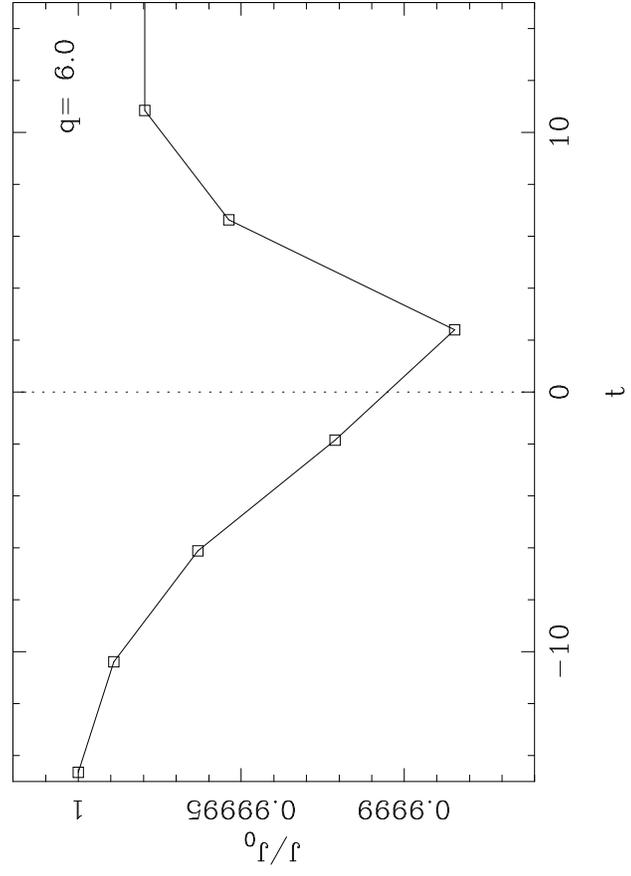
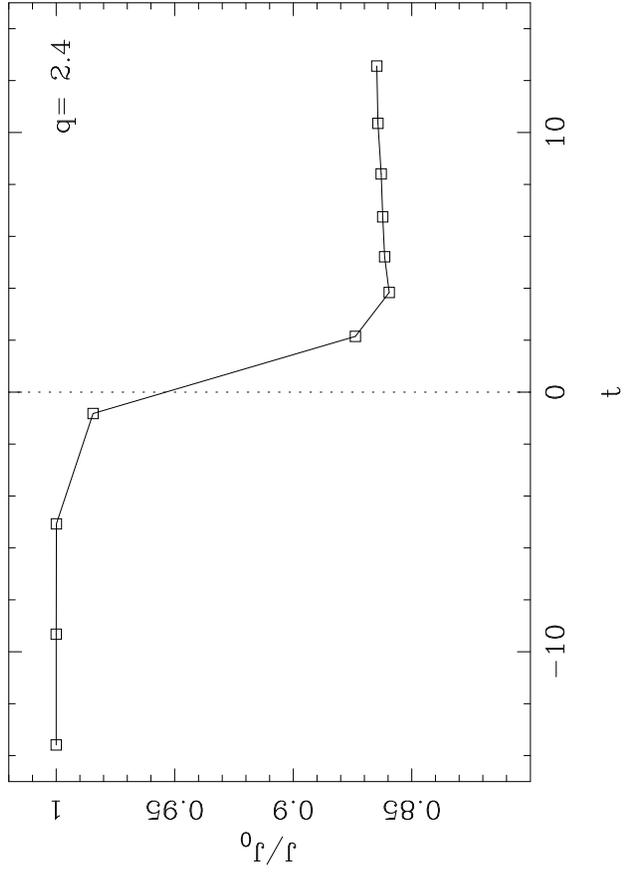
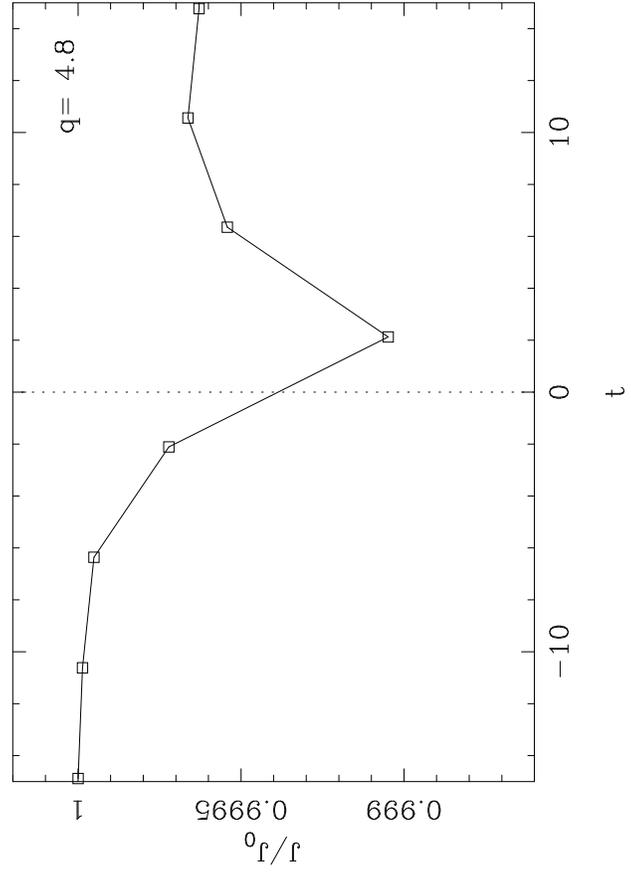
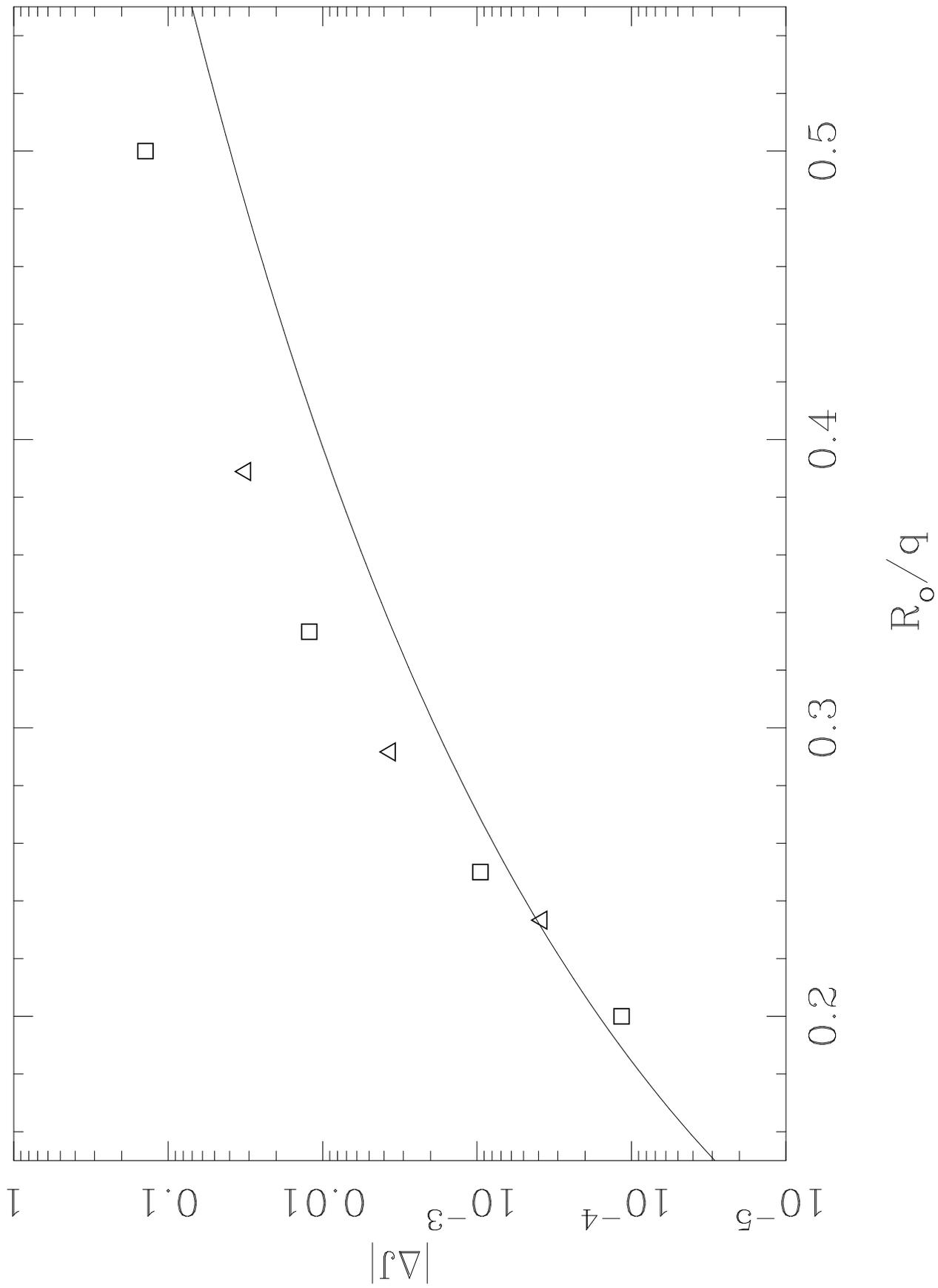
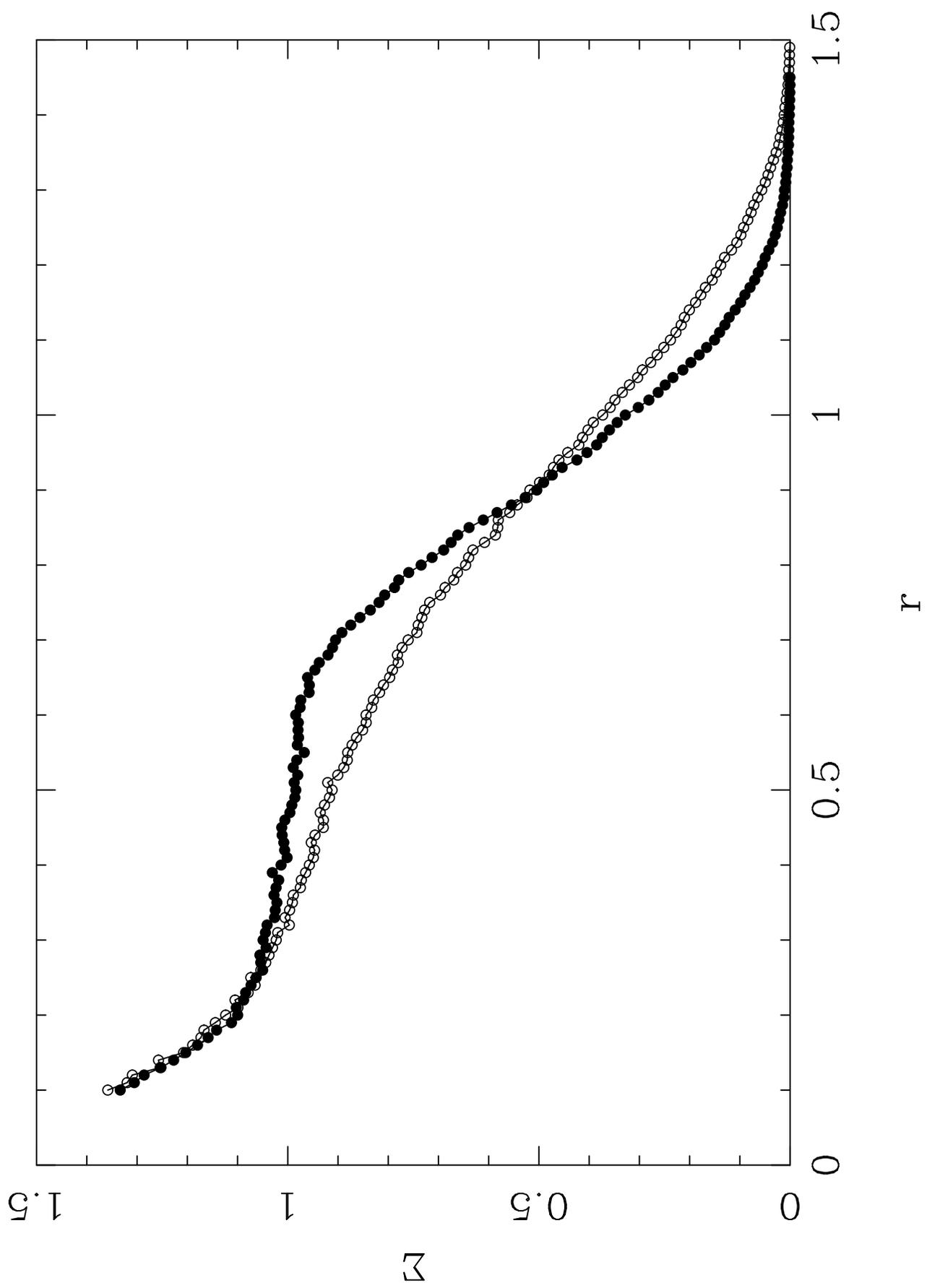
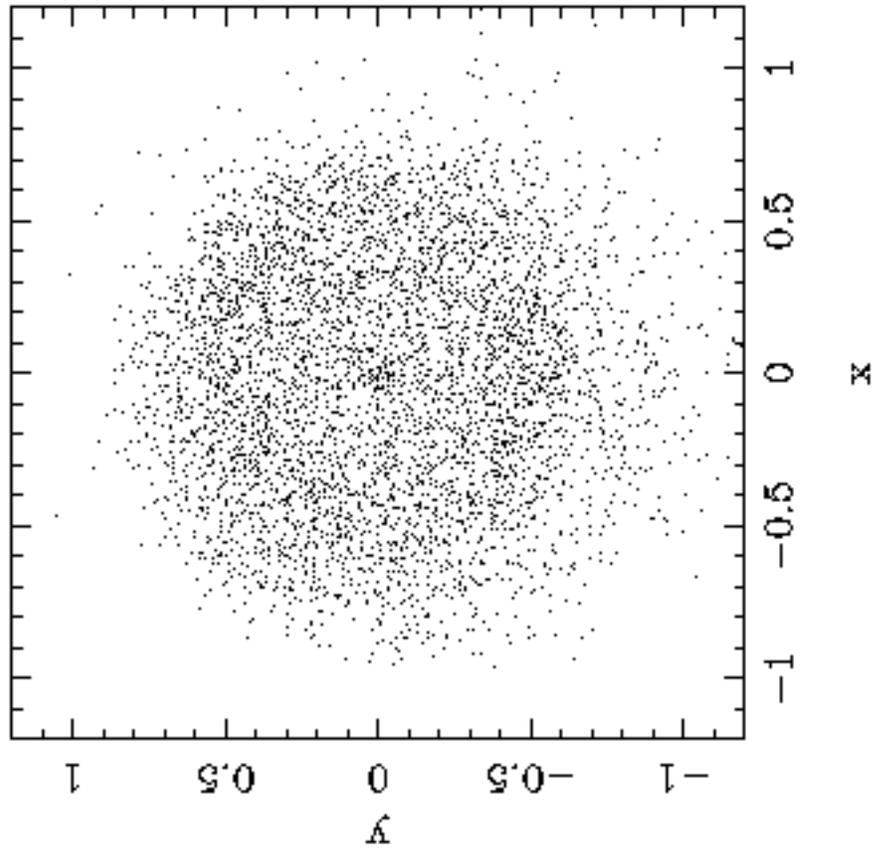
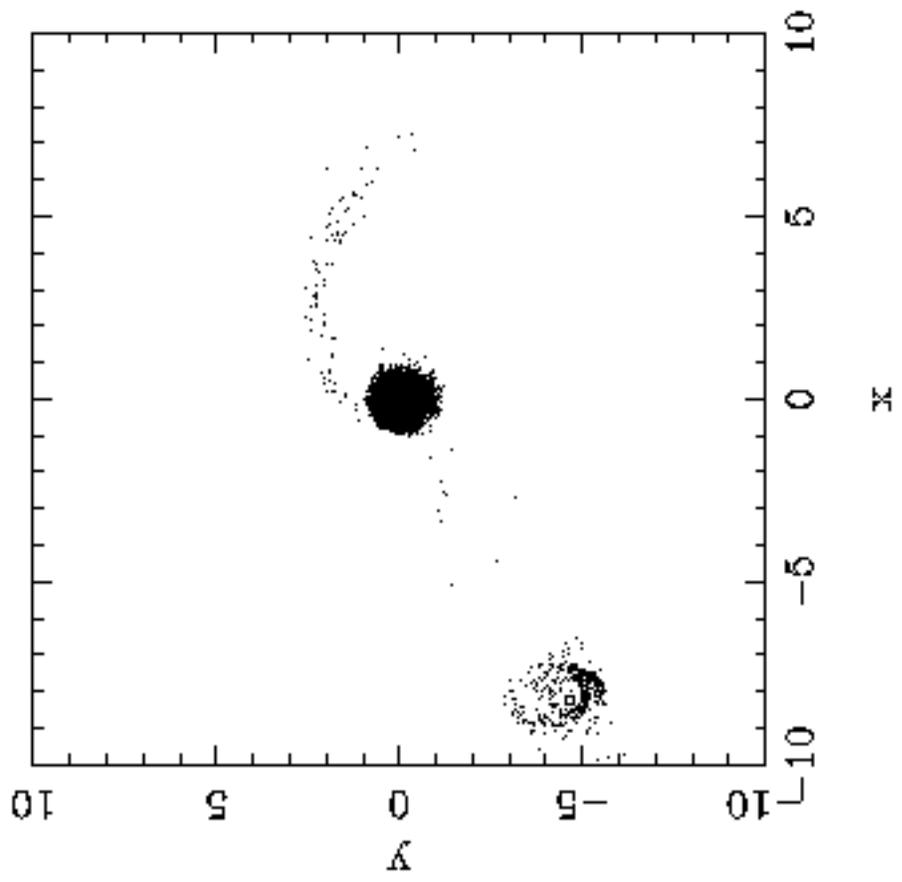
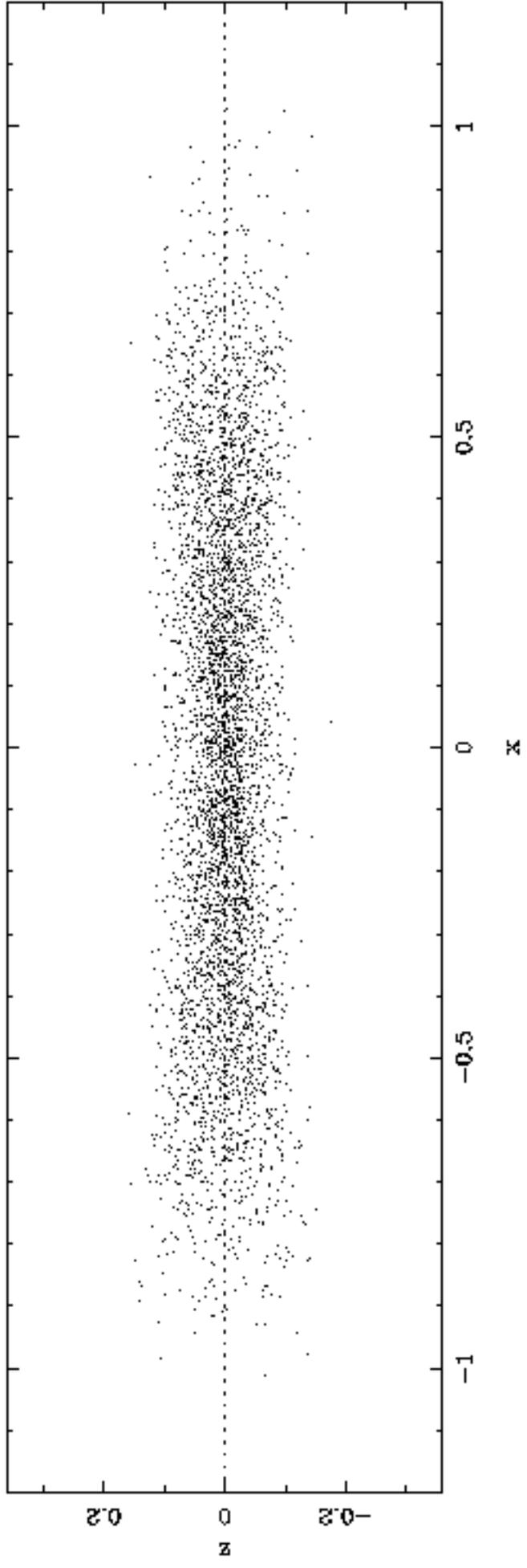
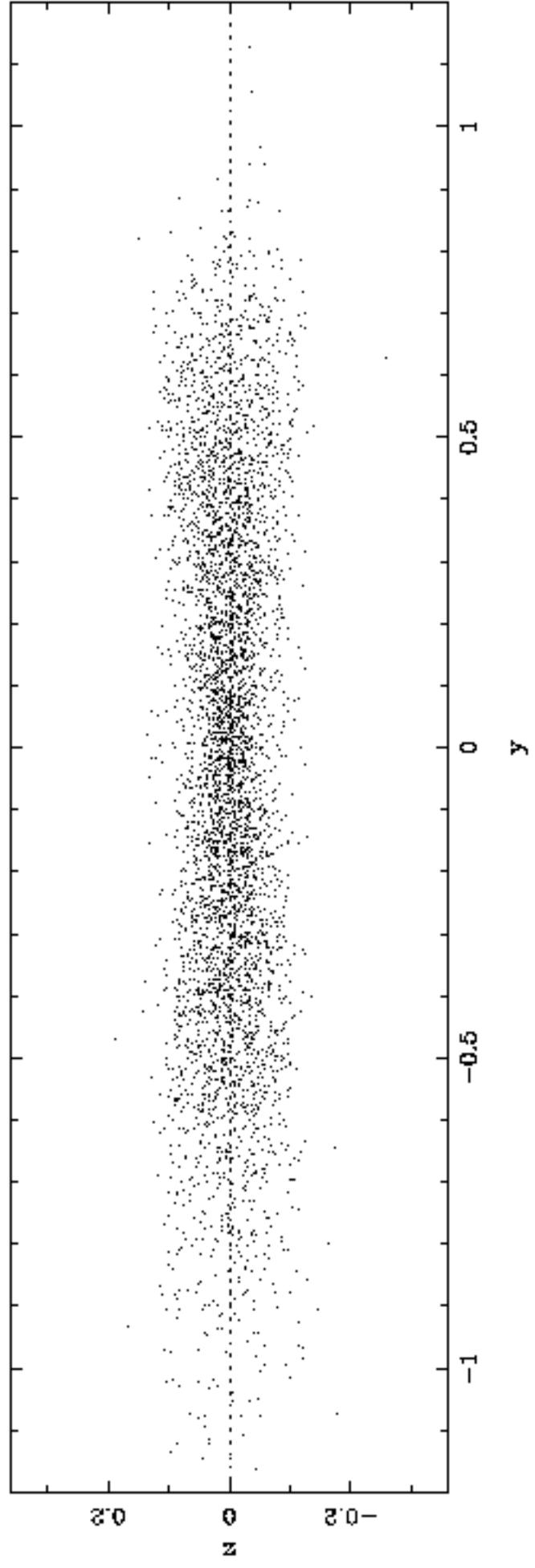
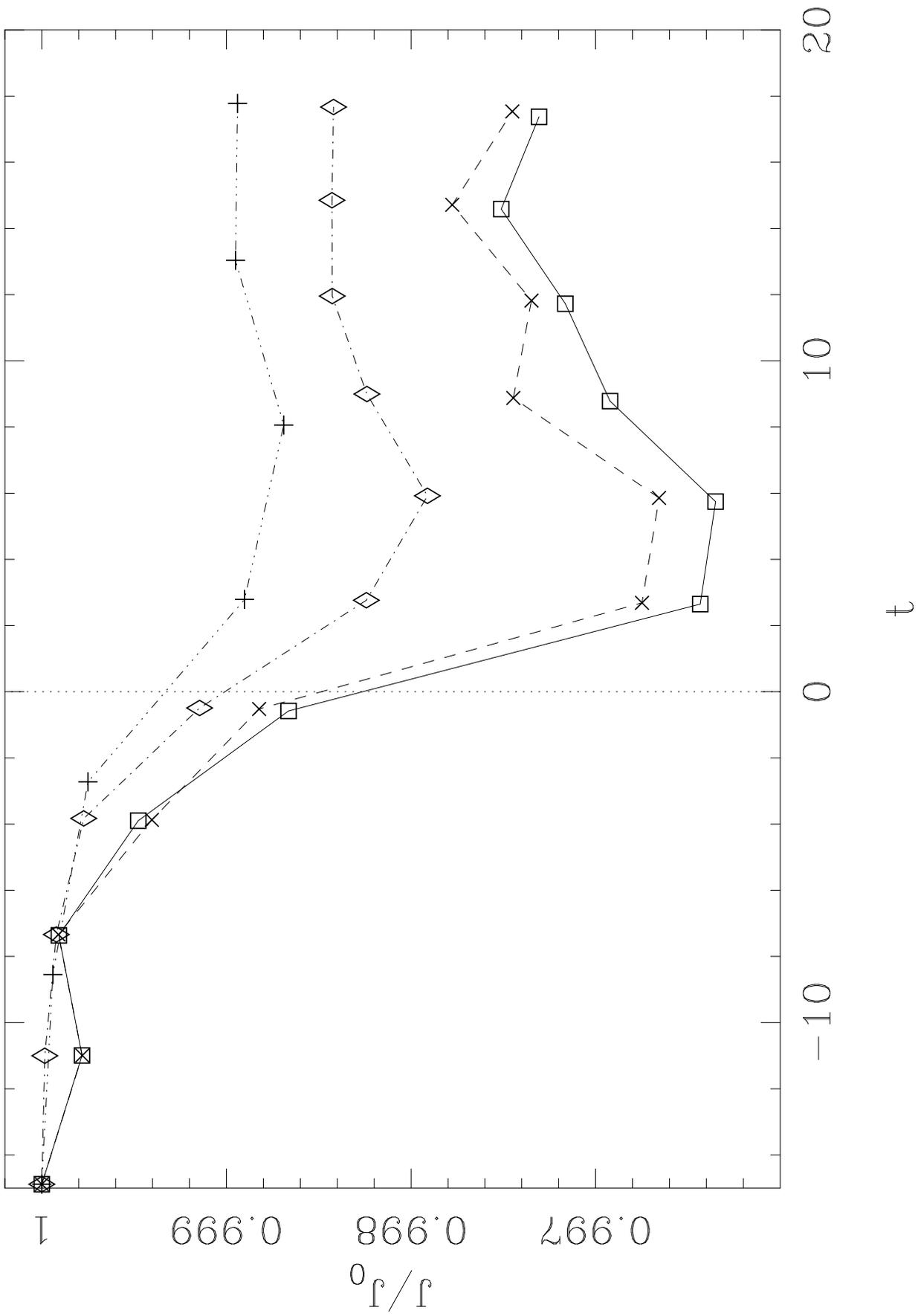
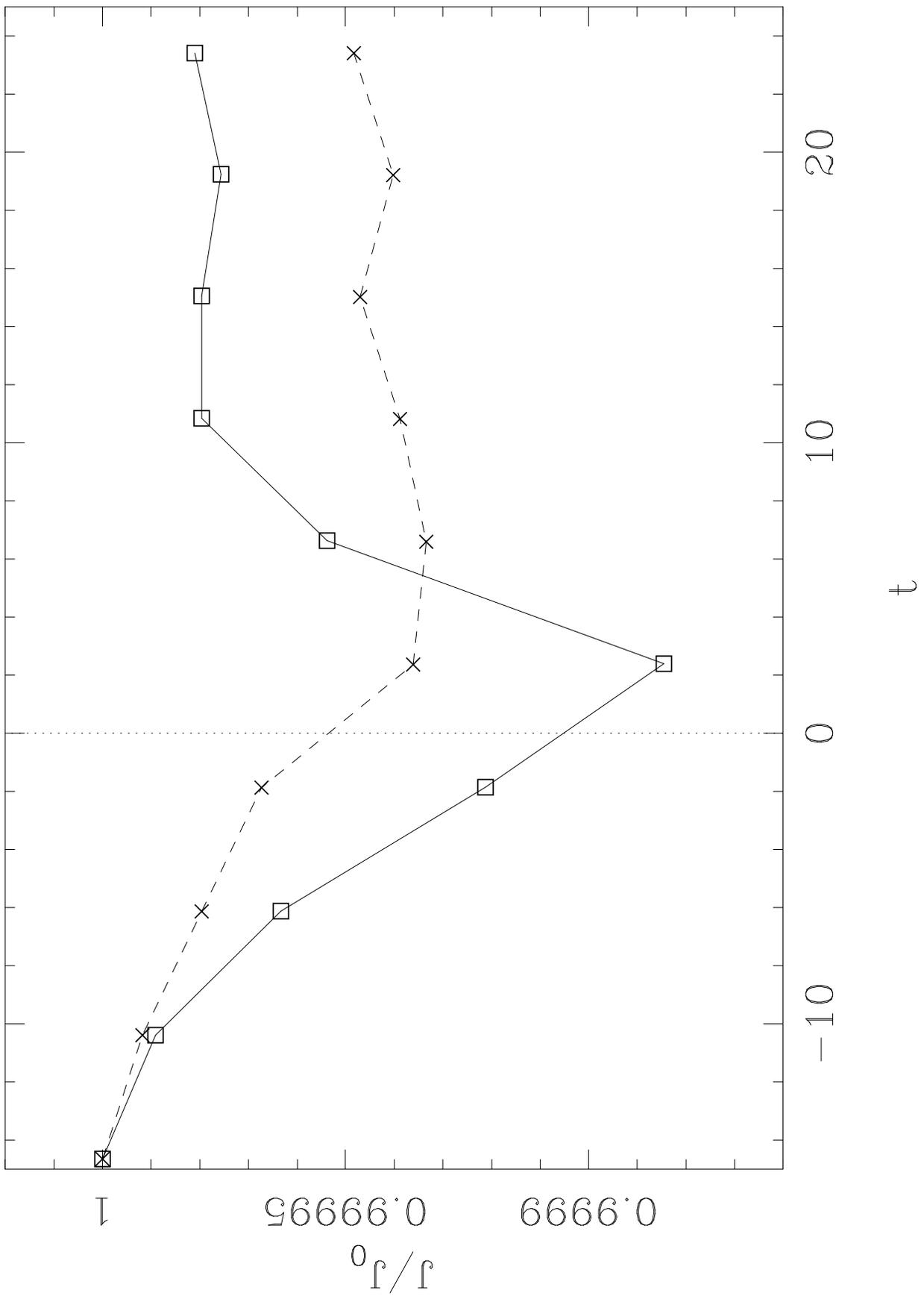
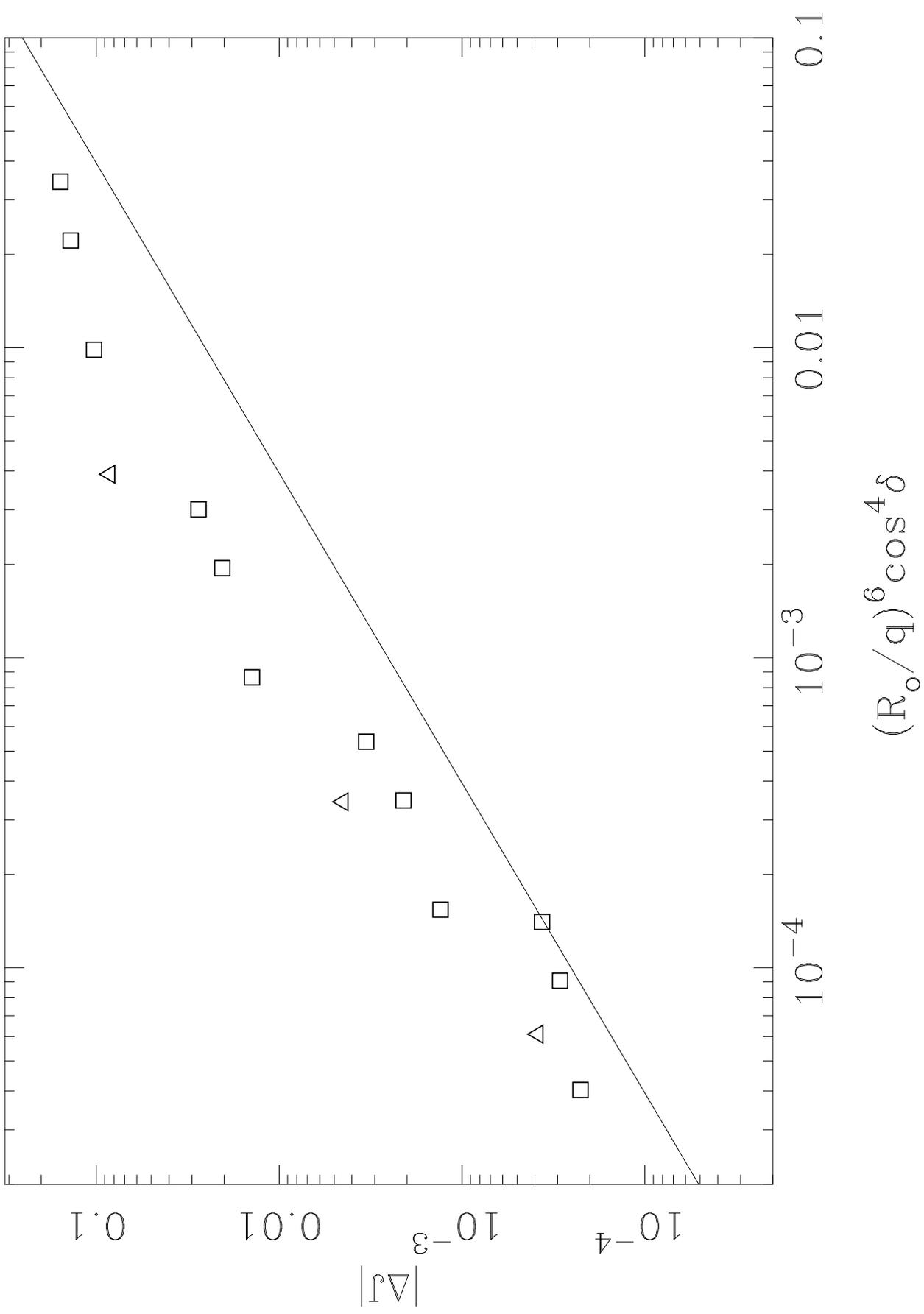
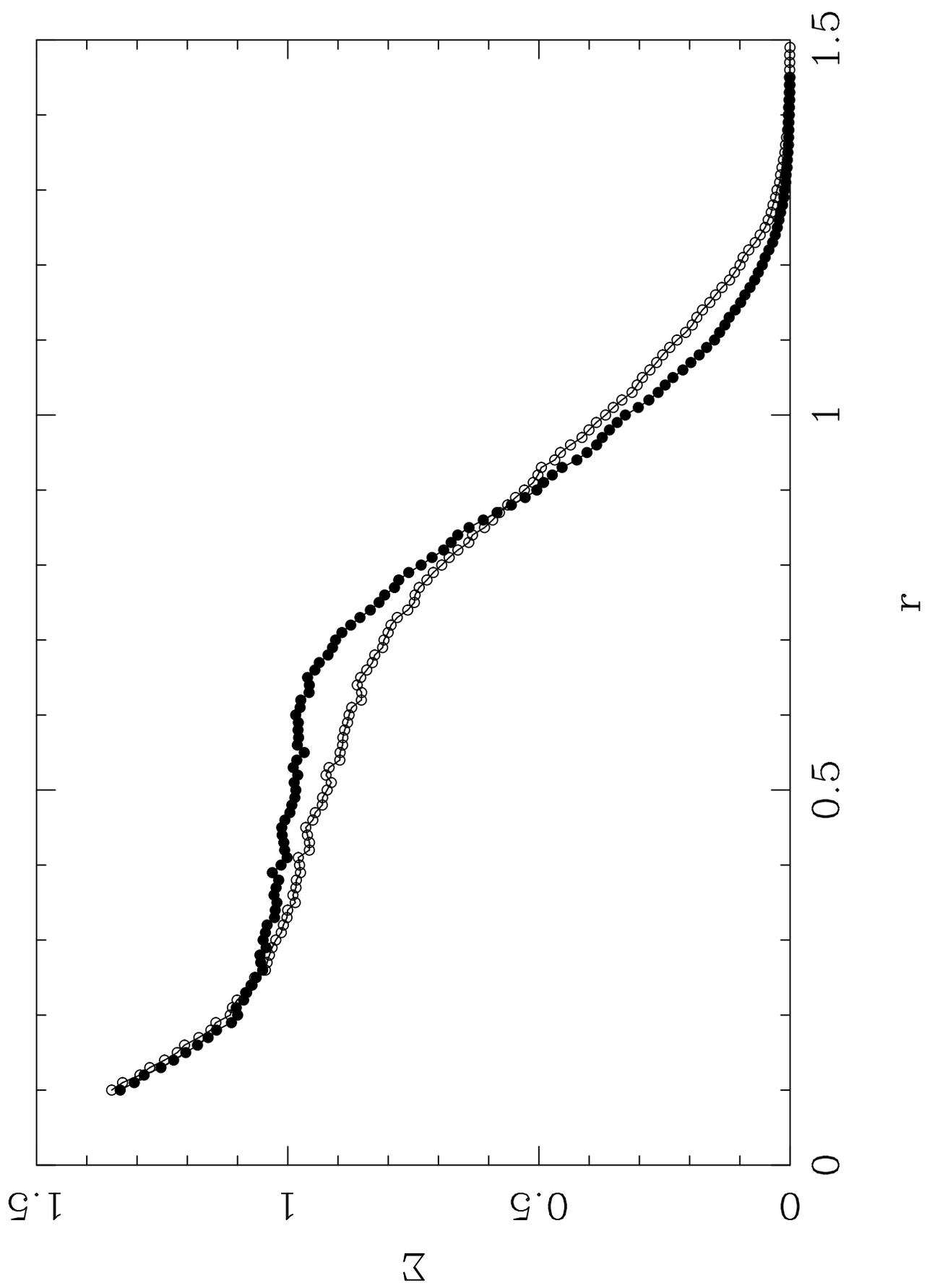
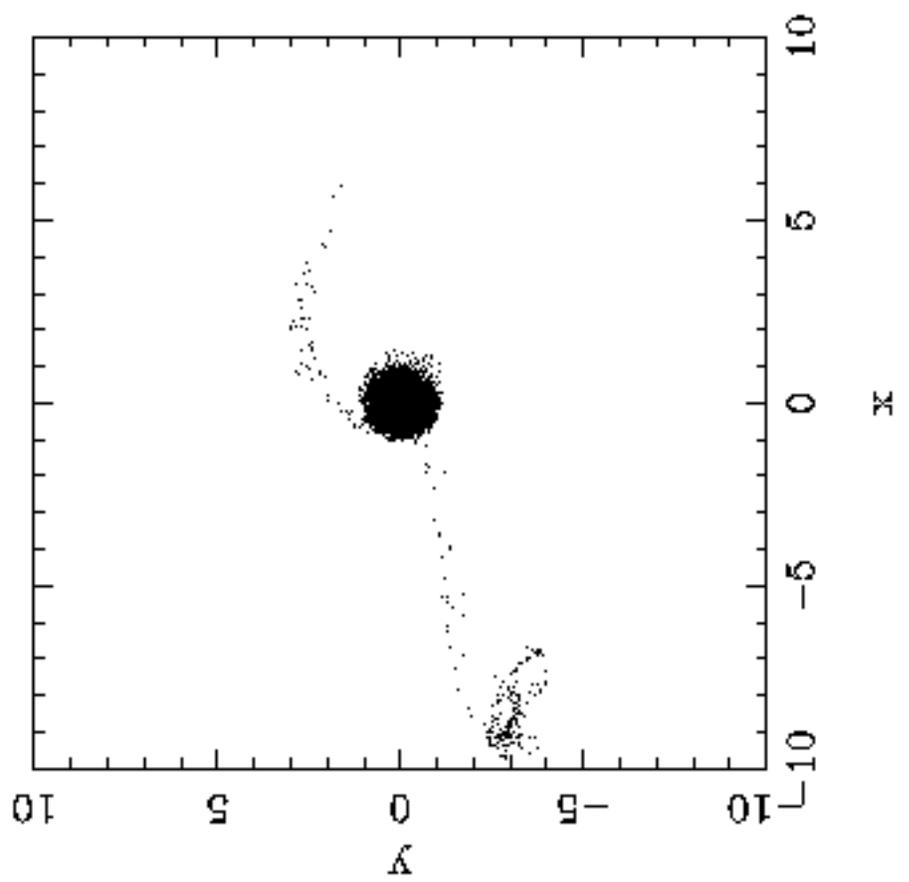
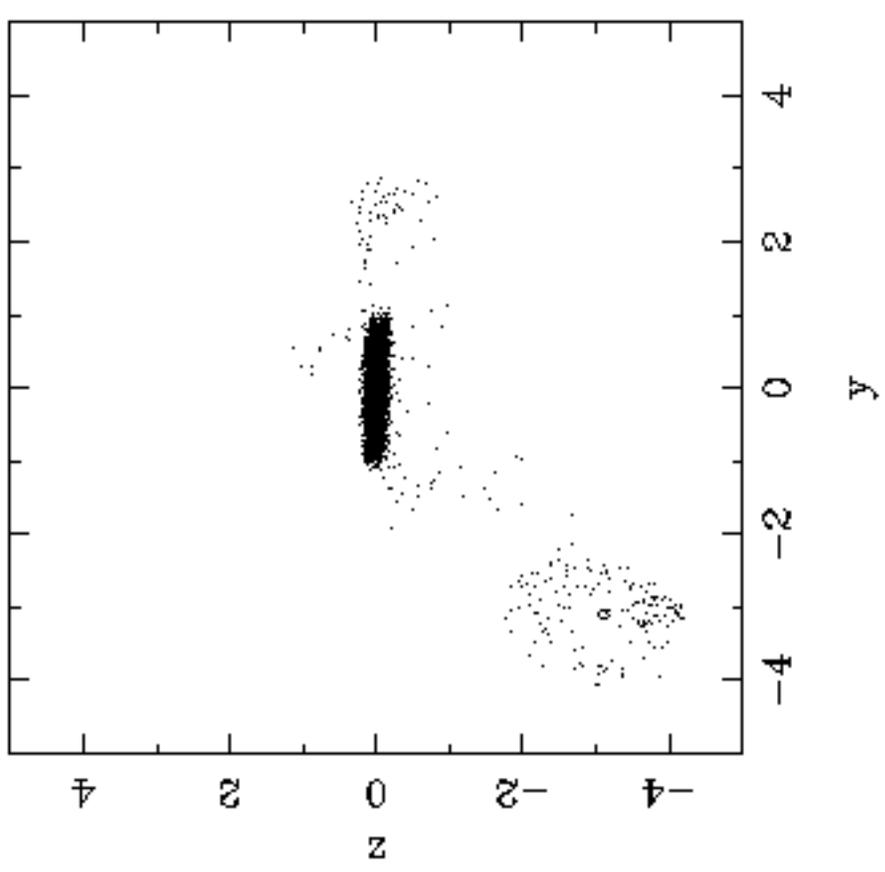
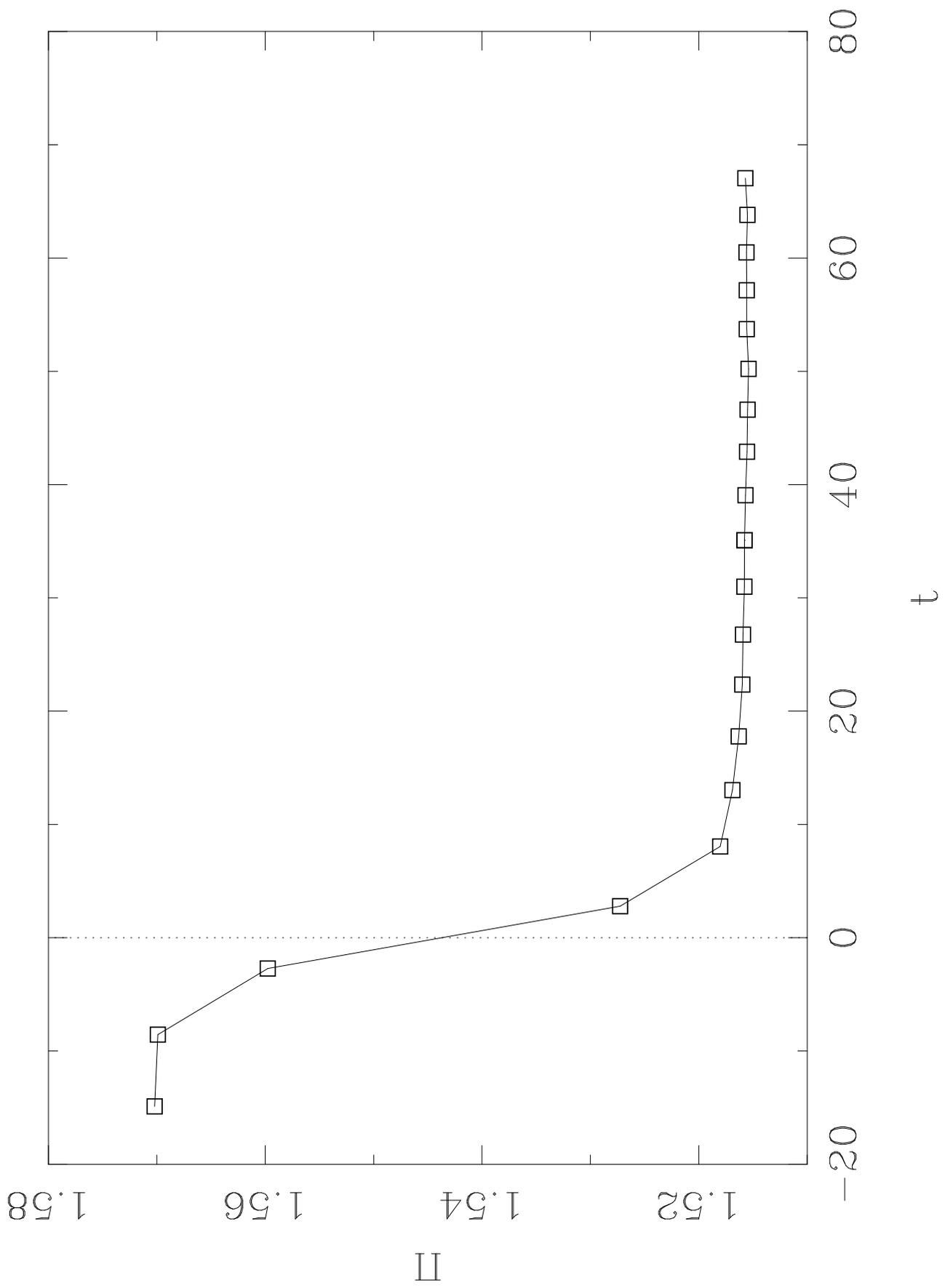
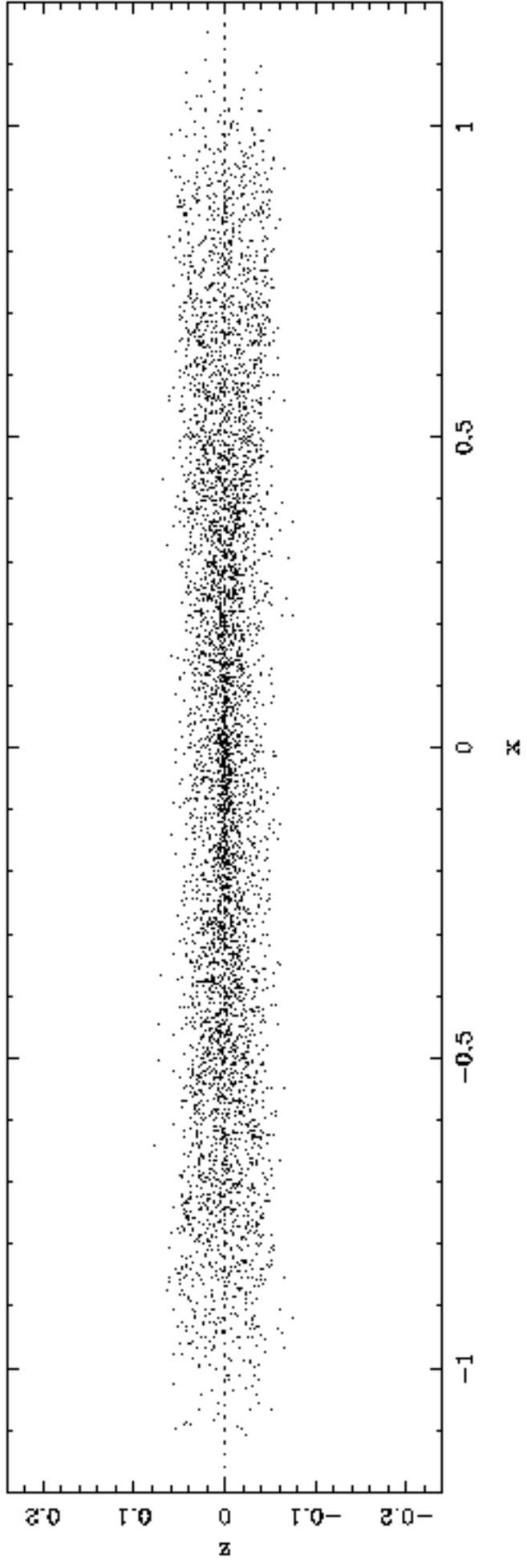
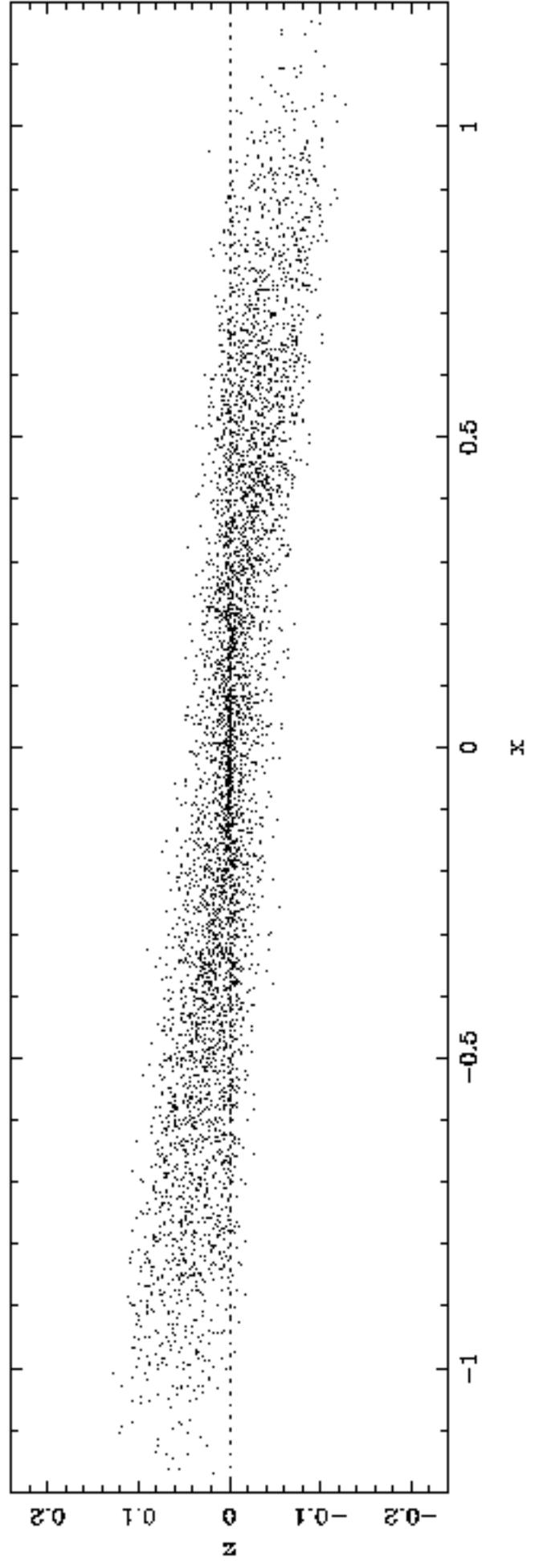
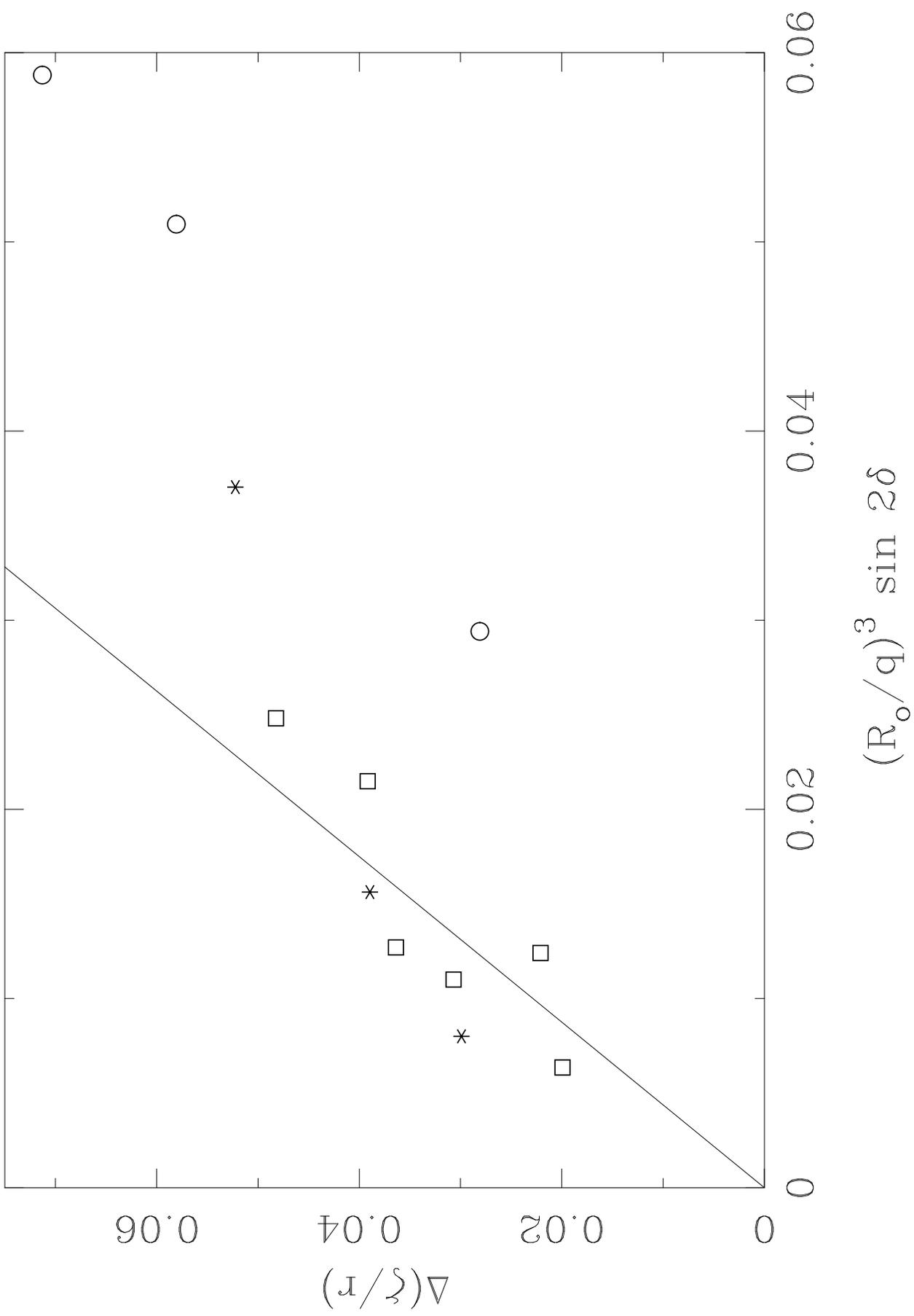
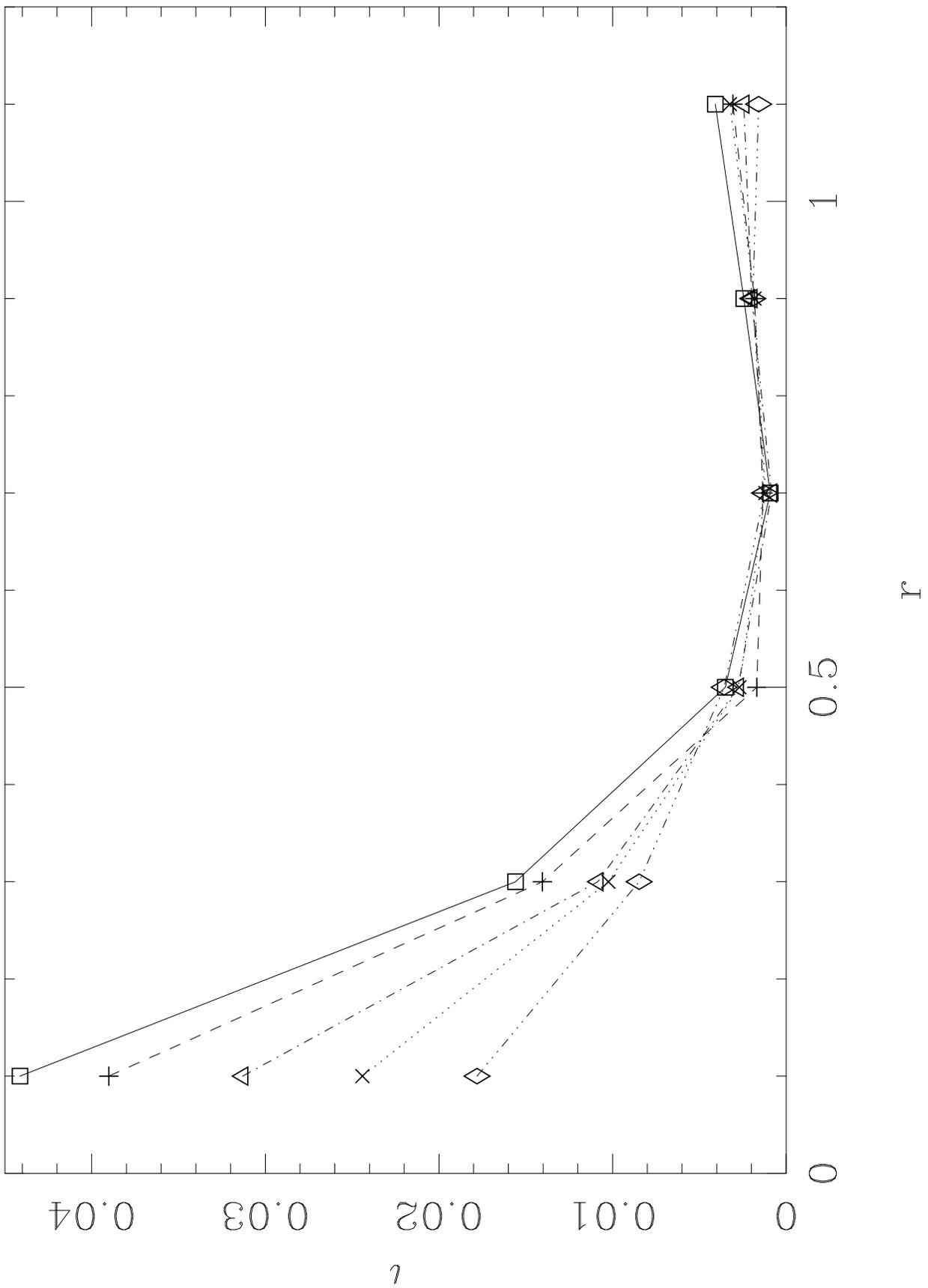
\end{document}

%% file: mnmacros.tex
\newif\ifAMStwofonts
\AMStwofontstrue

\ifoldfss
  \ifCUPmtlplainloaded \else
    \NewTextAlphabet{textbfit} {cmbxti10} {}
    \NewTextAlphabet{textbfss} {cmssbx10} {}
    \NewMathAlphabet{mathbfit} {cmbxti10} {} 
    \NewMathAlphabet{mathbfss} {cmssbx10} {} 
  \fi
  \ifAMStwofonts
    \ifCUPmtlplainloaded \else
      \NewSymbolFont{upmath} {eurm10}
      \NewSymbolFont{AMSa} {msam10}
      \NewMathSymbol{\uDelta}  {0}{upmath}{1}
      \NewMathSymbol{\ubeta}   {0}{upmath}{C}
      \NewMathSymbol{\udelta}  {0}{upmath}{E}
      \NewMathSymbol{\upi}     {0}{upmath}{19}
      \NewMathSymbol{\umu}     {0}{upmath}{16}
      \NewMathSymbol{\upartial}{0}{upmath}{40}
      \NewMathSymbol{\leqslant}{3}{AMSa}{36}
      \NewMathSymbol{\geqslant}{3}{AMSa}{3E}

       \let\ge=\geqslant
    \fi
  \fi
\fi 

\ifnfssone
  \newmathalphabet{\mathit}
  \addtoversion{normal}{\mathit}{cmr}{m}{it}
  \addtoversion{bold}{\mathit}{cmr}{bx}{it}
  \newmathalphabet{\mathbfit} 
  \addtoversion{normal}{\mathbfit}{cmr}{bx}{it}
  \addtoversion{bold}{\mathbfit}{cmr}{bx}{it}
  \newmathalphabet{\mathbfss} 
  \addtoversion{normal}{\mathbfss}{cmss}{bx}{n}
  \addtoversion{bold}{\mathbfss}{cmss}{bx}{n}
  \ifAMStwofonts
    \ifCUPmtlplainloaded \else
      \UseAMStwoboldmath
      \makeatletter
      \new@mathgroup\upmath@group
      \define@mathgroup\mv@normal\upmath@group{eur}{m}{n}
      \define@mathgroup\mv@bold\upmath@group{eur}{b}{n}
      \edef\UPM{\hexnumber\upmath@group}
      \new@mathgroup\amsa@group
      \define@mathgroup\mv@normal\amsa@group{msa}{m}{n}
      \define@mathgroup\mv@bold\amsa@group{msa}{m}{n}
      \edef\AMSa{\hexnumber\amsa@group}
      \makeatother
      \mathchardef\uDelta="0\UPM1
      \mathchardef\ubeta="0\UPMC
      \mathchardef\udelta="0\UPME
      \mathchardef\upi="0\UPM19
      \mathchardef\umu="0\UPM16
      \mathchardef\upartial="0\UPM40
      \mathchardef\leqslant="3\AMSa36
      \mathchardef\geqslant="3\AMSa3E

       \let\ge=\geqslant
    \fi
  \fi
\fi 

\ifnfsstwo
  \DeclareMathAlphabet{\mathbfit}{OT1}{cmr}{bx}{it}
  \SetMathAlphabet\mathbfit{bold}{OT1}{cmr}{bx}{it}
  \DeclareMathAlphabet{\mathbfss}{OT1}{cmss}{bx}{n}
  \SetMathAlphabet\mathbfss{bold}{OT1}{cmss}{bx}{n}
  \ifAMStwofonts
    \ifCUPmtlplainloaded \else
      \DeclareSymbolFont{UPM}{U}{eur}{m}{n}
      \SetSymbolFont{UPM}{bold}{U}{eur}{b}{n}
      \DeclareSymbolFont{AMSa}{U}{msa}{m}{n}
      \DeclareMathSymbol{\uDelta}{0}{UPM}{"1}
      \DeclareMathSymbol{\ubeta}{0}{UPM}{"C}
      \DeclareMathSymbol{\udelta}{0}{UPM}{"E}
      \DeclareMathSymbol{\upi}{0}{UPM}{"19}
      \DeclareMathSymbol{\umu}{0}{UPM}{"16}
      \DeclareMathSymbol{\upartial}{0}{UPM}{"40}
      \DeclareMathSymbol{\leqslant}{3}{AMSa}{"36}
      \DeclareMathSymbol{\geqslant}{3}{AMSa}{"3E}

       \let\ge=\geqslant
    \fi
  \fi
\fi 

\ifCUPmtlplainloaded \else
  \ifAMStwofonts \else 
    \def\uDelta{\Delta}
    \def\ubeta{\beta}
    \def\udelta{\delta}
    \def\upi{\pi}
    \def\umu{\mu}
    \def\upartial{\partial}
  \fi
\fi